\documentclass[a4]{article}                      
\usepackage{epsfig}
\begin{document}
{\rm

\title{Determination of Matter Surface Distribution \\
of Neutron-rich Nuclei}

\author{Akihisa KOHAMA $^1$
\thanks{Permanent Address: RI Beam Science Laboratory, RIKEN.
        e-mail:~ kohama@rarfaxp.riken.go.jp}, 
Ryoichi SEKI $^{2, 3}$, Akito ARIMA $^4$, \\ and Shuhei YAMAJI $^1$ \\
$^1$ {\normalsize RIBF Project Office, Cyclotron Center}, \\
     {\normalsize RIKEN (The Institute of Physical and Chemical Research)}, \\
     {\normalsize 2-1 Hirosawa, Wako-shi, Saitama 351-0198, JAPAN} \\
$^2$ {\normalsize Department of Physics and Astronomy}, \\
     {\normalsize California State University, Northridge, CA 91330, USA} \\
$^3$ {\normalsize W. K. Kellogg Radiation Laboratory}, \\
     {\normalsize California Institute of Technology}, 
     {\normalsize Pasadena, CA 91125, USA}  \\
$^4$ {\normalsize The House of Councilors}, 
     {\normalsize 2-1-1 Nagata-cho, Chiyoda-ku, Tokyo 100-8962, JAPAN} }
     
\maketitle

\begin {abstract}
We demonstrate that the matter density distribution in the surface region
is determined well by the use of the relatively low-intensity
beams that become available at the upcoming radioactive beam facilities.
Following the method used in the analyses of electron scattering, 
we examine how well
the density distribution is determined in a model-independent way
by generating pseudo data and by carefully applying statistical and 
systematic error analyses.
We also study how the determination becomes deteriorated in the
central region of the density, as the quality of data decreases.
Determination of the density distributions of neutron-rich nuclei is 
performed by fixing parameters in the basis functions
to the neighboring stable nuclei.  
The procedure allows that the knowledge of 
the density distributions of stable nuclei assists to 
strengthen the determination of their unstable isotopes.
\end{abstract}
{\small
\begin{flushleft}
\end{flushleft}
}

\newpage
\setcounter{section}{0}
\setcounter{equation}{0}
\section{Introduction}
\label{sec-intro}

Experimental facilities for unstable nuclei are receiving much attention
in the recent years.  Radioactive Ion Beam Factory (RIBF) is under
construction at RIKEN in Japan, and Rare Isotope Accelerator (RIA) facility
is being proposed in USA.
The facilities are expected to provide much data which are expected
to reveal interesting properties and dynamics of unstable
nuclei \cite{RIBF,RIA}.  Among various properties, one property of interest
is the one-body matter density distribution. 
The density distribution is a fundamental quantity of nuclei and 
serves as an important measure to test how well we understand 
nuclear structure \cite{ANT:OX}, and will be experimentally investigated
at an early stage after completion of the facilities.
It is perhaps an appropriate time to make a close investigation
how well one could determine the one-body matter density 
distributions from the data expected to emerge from the facilities.
As a concrete example, we examine the type of the experiments under
consideration at the RIBF.

Unstable nuclei of neutron-rich side (neutron-rich nuclei)
are often characterized as those with a large surface
region generated by loosely bound valance
nucleons \cite{Tani,Niigata,Ber:PR,Tani:JPG}. 
We will give a special attention to the determination 
of the matter density distributions in the surface
region, though we will have to treat, of course, the entire region of nuclei.  
The surface is the region where $\rho(r)$ $/ \rho(0)$ varies from 0.9 to 0.1 
if we have a density distribution of monotonically decreasing 
function in $r$, such as the Woods-Saxon density, in mind \cite{Elt}.
More elaborated definitions of the surface can be found 
in the literature \cite{BM:NPA}.

In this work, we focus on the question, 
how the precision in the determination of the matter density
distributions of neutron-rich nuclei 
is related to the accuracy of experimental data, specifically
of proton-nucleus elastic scattering differential cross sections. 
For the purposes of making the examination definitive, we 
perform a model calculation, by artificially generating data from given
density distributions and then by making their statistical analysis.

The question which we address is an old problem, and  
the basic method that we use has been well developed for extracting nuclear
charge density distributions from high-energy electron scatterings and muonic atom
data over years \cite{Bat:ANP,Sic:LNP,Fro:MT,FN:NP}.  We make no new
contribution to the method, but apply it to the determination of
the nuclear matter density distributions, focusing on the surface region of
neutron-rich nuclei.  As we discuss in the next section, the method has
been applied to the determination of the nuclear matter density distributions but
with less rigor, and no use has been made to the case of neutron-rich
nuclei.  The present work is novel in the careful examination of how the
method is applied to the nuclear-matter-density-distribution problem.

In this work, we follow the traditional approach 
of the least-square method \cite{Cowan}
and do not discuss other approaches often referred to as Bayesian methods,  
such as maximum entropy technique.  These approaches involve issues
of subjectivity \cite{Cowan,MEM1}, and require a separate study. 
We are currently investigating their relevance to our problem.

The contents of this paper are as follows:
We describe our approach in comparison to previous works in
Section~\ref{sec-over}, and the formalism used in our analysis in 
Section~\ref{sec-formalism}.
There, we introduce pseudo data in Subsection~\ref{sec-pdata} and
error estimate in Subsection~\ref{sec-estim}.
Numerical results are presented in Section~\ref{sec-nume},
including discussions of pseudo data in Subsection~\ref{sec-numpd} and
those of error estimate in Subsections~\ref{sec-numcom} and \ref{sec-fit}.
We summarize our results in Section~\ref{sec-summary}. 
Appendices include a brief discussion of basis functions
in Appendix~\ref{sec-kgauss}, a list of the expressions for the density 
and the data in Appendix~\ref{sec-xsec}, and a summary of formula and
numerical results concerning the Fourier-Bessel basis functions
in Appendix~\ref{sec-fb}.

Some preliminary results of this work have been previously 
reported \cite{drip}.  This paper provides a full account of our work 
\footnote{This paper is the revised version of the preprint, 
RIKEN-AF-NP-413, Nov., 2001, with the same authors.}.
The major added contents are the discussion of the completeness error, 
the way how to generate pseudo data, 
and the other confirmation of the fitted results 
with the use of the $\chi^2$-distribution. 
We have also added several Appendices to the previous report.

\newpage
\setcounter{equation}{0}
\section{Overview of Our Approach}
\label{sec-over}

In this investigation, the density distributions are determined 
in a ``model-independent" way.  As the model-independent determination 
is the central feature of our procedure, we wish to clarify at the outset
what we mean by the model-independent procedure.   
It is model-independent, 
because the density distribution is determined 
as a linear combination of basis functions. 
We do this by avoiding the use of specific analytic forms 
of the density distribution, which is the common practice in this field 
\cite{Bat:ANP,Sic:LNP,Fro:MT,FN:NP}. 

We will explicitly 
assume, however, that the density distribution is directly related to the 
differential cross sections through the eikonal approximation 
in the lowest-order optical potential, so-called $t$-$\rho$ form \cite{Glau:Lec}.  
We assume this because of convenience and simplicity.  
The assumption is basically 
independent of our semi-quantitative conclusion
on the question of how the density determination depends on experimental accuracy.  
That is, we expect that the same conclusion would emerge if we were to adopt 
a more elaborate relation.  We emphasize that the issue we wish to address 
in this paper is this question of the dependence of the density 
{\it determination} on the experimental {\it accuracy}, not on the relation 
between the density and the cross section themselves. 

The elastic scattering cross section data that we will examine 
in our model calculation are generated artificially, 
and will be referred to as ``pseudo data" 
\cite{FN:NP,Dre:NPA,FV:NPA,FVR:NPA}.  
At first, they are generated from a specific density distribution 
(of the three-parameter Fermi form, Eq.~(\ref{eq:numcom1})) 
by the use of the above described procedure with the eikonal approximation 
and are then shuffled statistically 
at each data point with an assigned 
experimental uncertainty (corresponds to a Gaussian width), 
simulating the situations expected in future experiments at the RIBF. 
We will prepare 25 sets of such shuffled data, each of which yields 
the best-fit density distribution with its uncertainty.
The originally assumed density distribution is taken to be the true 
density distribution.  The average of the density distributions determined 
will be close to the true distribution, and their deviations from 
the average will give us the statistical uncertainty associated with the 
model-independent determination \cite{Cowan}.  The difference between the true 
and averaged distributions corresponds to the systematic uncertainty \cite{Cowan}.  
The systematic uncertainty corresponds to the limitation of 
the model-independent procedure associated with our (model-independent) 
representation of the density distribution.  The systematic uncertainty 
could also include the uncertainty associated with the relation between the 
density distribution and the differential cross sections.  In this work, 
we will not examine this part of the systematic uncertainty by applying the 
same relation for generating the data and fitting to them. 

We apply a particular combination of Gaussian functions \cite{Kam:PRA} 
as a model-independent representation of the density distribution.  
We do this for simplicity and for stability of computation, 
but a particular choice
of the basis should not alter the final conclusion.  That is, the nuclear
density distribution determined at the end should fall in within the
systematic error that is noted above.  
We will confirm this by comparing the results of our Gaussian basis functions 
with those of the more established Fourier-Bessel basis \cite{FN:NP}.  

There are several well-known basis functions. 
The expansion in terms of $\delta$-functions 
was proposed originally by Lenz \cite{Lenz:ZP}, 
and extended by Lenz and Friedrich \cite{FL:NP}. 
Since the early 70's, Sick and his collaborators have been analyzing 
the electron scattering data 
using another set of Gaussian basis functions, 
the expansion in a sum of Gaussians (SOG) \cite{Sic:NP}, 
with great success \cite{Fro:MT,Bre:PRL,Cave:PRL}. 
These Gaussian basis functions are a series of Gaussians of multi-center. 
With their analyses, they demonstrated that the determinations of 
the charge density distributions by the phase-shift analyses 
is reliable and accurate by comparing the electron-nucleus 
with the positron-nucleus elastic scattering \cite{Bre:PRL}, 
and the charge density distributions of $^{208}$Pb was 
determined with an uncertainty of 
the order of $\Delta \rho(r)/\rho(r)$ $= \pm 1~[\%]$ \cite{Cave:PRL}.
Recenty, Burov {\it et al.} analyzed the electron scattering data 
by expanding the density distribution in terms of 
the symmetric Fermi distribution \cite{Eld:SJNP}
and its derivative series \cite{Bur:PAN}. 

The three-parameter Fermi distribution that we use mimics the matter density
distribution of $^{58}$Ni 
and the relativistic mean field result mimics that of $^{78}$Ni, 
which we chose as examples of stable and unstable nuclei, respectively. 
We assume here that the matter density distribution of $^{58}$Ni 
behaves similarly to its charge density.  From the intense studies 
of the charge density distribution \cite{Fic:PLB,Sic:PRL35}, 
it was found that the charge density of $^{58}$Ni shows 
the less oscillating structure 
than the predictions of Hartree-Fock calculations \cite{Sic:PRL35}. 
This validates our choice of this simple functional form
(See also in Appendix~\ref{sec-xsec}).
The pseudo data are generated for the proton incident energy of 1.047~[GeV].  
The energy is chosen to be the same as that of the 
proton-nucleus scattering experiments at SATURN, 
Saclay \cite{Alk:PLB,Alk:NPA,Lom:NPA}, 
because the experimental data and their analyses have shown that 
the eikonal approximation yields reasonably realistic results 
(See Fig.~1 in the next section).  

In this work, we assume that the proton-nucleus elastic scattering differential
cross sections and the nuclear density distribution are exactly related
by the eikonal approximation with the $t$-$\rho$ form. 
With this simple form, one can see analytically
that the nuclear radius determines the oscillation 
and the diffuseness does the exponential decrease with momentum transfer 
of the elastic cross section \cite{ADL80}. 
This helps one to obtain the insight to the relation 
between the quality of data and the fitted density distributions. 
Of course, the eikonal approximation is an approximation, 
and the degree of its validity is 
sometimes questioned, especially when it is used with the $t$-$\rho$ form.
It can be numerically improved by the use of the partial-wave
decomposition method, and the $t$-$\rho$ form can be also improved 
by incorporating higher-order effects into the potential.  
Our objective
of finding the relation between the precision of the density determination
and the experimental accuracy certainly depends on how reliably the density
and the data are related each other.  
For the sake of making our examination as clear as possible, however, 
we isolate 
the issue of the precision-accuracy in this model calculation, by artificially 
assuming the density-data relation to be exact.   In realistic
model-independent analyses of actual experimental data, we must address
the latter issue and take account of the reliability of the relation
of the two quantities.

Once the density-data relation is assumed to be exact, the model-independent 
analysis can provide a definitive result for the precision-accuracy issue 
at any energy.  The result will become, however, unrealistic as we go down 
in the energy.  For a model calculation at lower energies, such as those 
around 500~[MeV] where various experiments are being planned at the RIBF, 
it would be desirable to use more rigorous relation based on the 
partial-wave decomposition and an improved optical potential.  
We are planning to extend the present work to these energies 
by the use of partial-wave decomposition. 

There have been efforts on the determination of the nuclear (matter-)
density distributions \cite{Bat:ANP,AWT:NP81,Alk:PR,Cha:AP}, 
which are, of course, founded on the charge density studies. 
As reviewed by Alkhazov {\it et al.} \cite{Alk:PR}  
and by Chaumeaux {\it et al.} \cite{Cha:AP} in 1978, 
and by Batty {\it et al.} \cite{Bat:ANP} in 1989, 
some works include model-independent analyses of the experimental data: 
after the first attempt of the model-independent approach 
to the matter density distributions by Brissaud {\it et al.} \cite{PRC:Bri}, 
Ray {\it et al.} \cite{Ray:PRC,Ray:PRC79}, 
Hoffmann {\it et al.} \cite{Hof:PRC}, 
and Alkhazov {\it et al.} \cite{Alk:NPA} studied 
the matter density distributions along similar lines. 
More recent work is by Starodubsky {\it et al.} \cite{Str:PRC}.
Gils {\it et al.} analyzed the data 
of the $\alpha$-nucleus elastic scattering, 
and obtained the quite similar surface density distribution as those from 
the proton-nucleus scattering data
within the statistical uncertainties \cite{Gil:PRC}. 
None of these works has addressed, however, the issue of determination
of the surface density distribution associated with nuclei near the drip lines. 
The recent analysis by Alkhazov {\it et al.} \cite{Alk:PRL}
of the GSI experiment of helium isotope addresses this issue, but it
is a model-dependent analysis and its conclusion is in question
under the light of a model-independent analysis \cite{seki}.

Our work is the first model calculation that examines closely 
the information content of the experimental data, including
the question of the {\em systematic} error 
as well as the statistical error, in the model-independent approach.

Although they are model-dependent,  
the matter density distributions of neutron-rich nuclei 
have been deduced recently, {\it e.g.} from the low-energy proton elastic 
and inelastic scattering \cite{Mar:PRC,Sch:PRC}, 
and from the interaction cross sections \cite{Tani:PL,Ozawa1,Ozawa}.

\newpage
\setcounter{equation}{0}
\section{Formalism}
\label{sec-formalism}

\subsection{Fitting Procedure}
\label{sec-form}

For the sake of notations and completeness, we outline our formalism 
in this section.  It is basically the same as that in the review article 
by Friar and Negele \cite{FN:NP}. 
The formalism itself was studied also in detail by Dreher {\it et al.}
\cite{Dre:NPA}.

We approximately express the one-body matter density distribution 
that is assumed 
to be true, $\rho_{\rm true}({\bf r})$, 
in $M$-terms of basis functions, $f_n({\bf r})$, $(n = 1, \cdots, M)$:
\begin{eqnarray}
  \rho_{\rm true}({\bf r}) \simeq 
  \rho_M^{\rm fit}({\bf r}) \equiv \sum_{n = 1}^{M}  C_n \; f_n({\bf r}),  
\label{eq:form1}
\end{eqnarray}
where $\rho_M^{\rm fit}({\bf r})$ is the fitted density distribution. 
Here, as we do not use a particular analytical functional form of the 
density, our approach is model-independent.  
Strictly speaking, there is no completely model-independent approach, 
because the basis functions include parameters to be chosen 
and only a finite number ($M$) of the basis functions can be used.  
The systematic errors are generated 
due to this incomplete model-independence of the approach. 
We will describe this point fully later. 

For simplicity, we assume that the nuclei are not deformed and examine only 
the radial dependence of their distributions.  The deformation can be treated 
in a similar model-independent way, but will make the procedure more 
complicated.  The fitted density is normalized to the mass number, $A$, as
\begin{eqnarray}
  \int d{\bf r}\; \rho_M^{\rm fit}(r) = 
  4 \pi \int_{0}^{\infty} r^2 dr\; \rho_M^{\rm fit}(r) = A. 
\label{eq:form2}
\end{eqnarray}
The expansion coefficients, $\{C_n\}$, are determined by minimizing 
\begin{eqnarray}
\chi^2(\{C_n\})
  = \sum_{\alpha = 1}^{N_{\rm data}}  {1 \over \epsilon_{\alpha}^2}\;
    \left( {d\sigma_{\alpha}^{\rm data} \over d \Omega}
         - {d\sigma_{\alpha}^{\rm fit} \over d \Omega}(\{C_n\}) \right)^2,  
\label{eq:form3}
\end{eqnarray}
under the variations of the finite set \{$C_n$\}. 
Here, $d\sigma_\alpha / d\Omega$'s are the differential cross sections 
of proton-nucleus elastic scattering 
at the center-of-mass angle, $\theta_\alpha$. 
$N_{\rm data}$ is the number of data points, and 
$\epsilon_{\alpha}$ is the uncertainty or error associated with the 
data at $\theta_\alpha$.  
${d\sigma_{\alpha}^{\rm data} / d \Omega}$'s are 
the pseudo data that we generate, and 
${d\sigma_{\alpha}^{\rm fit} / d \Omega}(\{C_n\})$'s are those 
calculated from the fitted density distribution.  
We carry out all least-square 
minimization using MINUIT \cite{minuit} in the CERN Library. 

The dependence of ${d\sigma_{\alpha}^{\rm fit} / d \Omega}$ on $\{C_n\}$ 
involves complicated physics.  In this work, we use the first-order 
optical potential in the Glauber approximation, which establishes the 
relation between the nuclear density 
and the differential cross section \cite{Glau:Lec}.
The relation is well known, but we list some of basic formulae 
in the relation in Appendix~\ref{sec-xsec} for completeness and clarification.
As noted in Section~\ref{sec-over}, 
we assume the relation to be exact and generate pseudo data using it.  
No real experimental differential cross section 
datum enters our calculation and its analysis.  
As such, it is logically immaterial how reliable the relation itself is, 
but it becomes an issue how realistic or relevant to future experiments. 

Figure~1 compares the real experimental data 
of proton-$^{58}$Ni elastic scattering cross sections 
at 1.047~[GeV] from SATURN, Saclay \cite{Alk:PLB,Alk:NPA,Lom:NPA} 
with the calculation. 
Note that in our later model calculations, 
we will use the same nuclear density at the same kinetic energy 
as used here. 
Agreement between the data and the calculation is not too great, but is 
not too much off either.  The agreement could be improved by including 
the Coulomb interaction and also various higher-order corrections, such as 
the center-of-mass correction and nuclear correlation \cite{Glau:Lec,GM:NPB}.   
The improvement 
would reduce the unpleasant sharp dips in the differential cross 
sections~\cite{ADL80,czy:NPB} and is certainly desirable, but again, 
is not essential to our model calculations.

As the basis functions of Eq.~(\ref{eq:form1}), 
we use Gaussian basis functions with a cut off at $R$:
\begin{eqnarray}
  f_n(r) = \exp ( - r^2 / r_n^2) \; \theta(R - r),~~~(n = 1, \cdots, M), 
\label{eq:form4}
\end{eqnarray}
where $r_n$ is the size parameter, defined in the geometrical progression, 
$r_n = r_1 a^{n - 1}$ $(n = 1, \cdots, M)$ \cite{Kam:PRA}.
These Gaussian basis functions differ from those used previously 
by Sick~\cite{Sic:NP}, and will be referred to as 
``Kamimura-Gauss basis functions" after the author who used them  
extensively \cite{Kam:PRA}.  
We lists the basic formulae of the Kamimura-Gauss basis 
functions in Appendix~\ref{sec-kgauss}.

In Subsection~\ref{sec-numcom} we will show that the completeness error defined 
in Subsection~\ref{sec-estim} of the Kamimura-Gauss basis functions is very small, 
which is the same order of magnitude as that of the Fourier-Bessel ones 
shown in Appendix~\ref{sec-fb}.

\subsection{Pseudo data}
\label{sec-pdata}

We create the total 25 sets of the pseudo data.  
We choose the number of 25 as the minimum number 
for statistical discussions. 
Each set mimics
a series of experimental data obtained by an experiment, and consists of
the proton-nucleus elastic scattering cross sections at the     
center-of-mass scattering angles that are common to all pseudo-data set.
No Coulomb interaction is included, for simplicity.  

The application of the pseudo data to estimate the density distributions 
is not new \cite{FN:NP,Dre:NPA}. 
In this work we study 
how far we can access the matter density distributions 
with varying the nuclear beam intensity, having in mind the neutron-rich nuclei. 
This is new. 

At each scattering angle, $\theta_{\rm \alpha}$, we thus generate 25 cross
sections.  They are generated artificially, but as realistically as possible.  
The following steps are taken to achieve this objective.   
First, we calculate the ``true" cross section, 
${d\sigma_{\alpha}^{\rm true}} / {d \Omega}$, from the given nuclear 
density of a three-parameter Fermi distribution form, as described in 
Appendix~\ref{sec-xsec}.  Second, a set of 25 cross sections are generated, 
randomly distributed about each ``true" cross section. 
Third, an uncertainty is assigned to each datum generated.

Let us elaborate on these steps. 
In the second step, 25 cross sections are generated following
the Gaussian distribution,   
\begin{eqnarray}
   f(x; \mu, \sigma^2)  =  {1 \over \sqrt{2 \pi \sigma^2}} \;
                                \exp \{ {-(x - \mu)^2 \over 2 \sigma^2} \} 
\label{eq:pd1b}
\end{eqnarray}
where
\begin{eqnarray}
  \mu    &=& {d\sigma_{\alpha}^{\rm true} \over d \Omega}, \\
  \sigma &=& \mu / \sqrt{N_{\rm yield}(\theta_{\alpha})}.
\label{eq:pd1}
\end{eqnarray}
Here, we define the yield count,
\begin{equation}
  N_{\rm yield}(\theta_{\alpha})
     =  {N_0 \over B_0} {d\sigma_{\alpha}^{\rm true} \over d \Omega},
\label{eq:pd2}
\end{equation}
where $B_0/N_0$~[mb/str] is the yield parameter, which is the magnitude 
of the cross section that yields a unit count in the measurement. 
We set $N_0 = 10$, so that
\begin{equation}
  \sigma = 0.316 \mu, ~~~ \mbox{\rm when} ~~~
  {d\sigma_{\alpha}^{\rm true} \over d \Omega} = B_0, 
\end{equation}
and we examine three cases of $B_0 =$ 0.1, 1.0, and 10.0 [mb/str]. 
Note that the datum less than $B_0 / N_0$ is rejected,
because $N_{\rm yield}(\theta_{\alpha})$ of 
Eq.~(\ref{eq:pd2}) becomes less than one. 

In the third step, we assign an uncertainty to each pseudo datum, 
using the expression of $\sigma$ in Eq.~(\ref{eq:pd1}) with
Eq.~(\ref{eq:pd2}), except for $d\sigma_{\alpha}^{\rm true} / d \Omega$
being replaced by the pseudo datum itself. 

The center-of-mass scattering angle is taken to be greater than
$4^{\circ}$, based on the expected experimental setups (of the inverse
kinematics) at the RIBF.  
The set of the angles, \{$\theta_{\rm \alpha}$\},
is chosen with $1.0^{\circ}$ separations, but the separation is reduced
near the forward direction as described in Appendix~\ref{sec-small}.
The number of the angles (and thus the number of the data in each 
pseudo-data set, $N_{\rm data}$) is then made to be more than 20, 
which provides the number of degrees of freedom, 
$N_{dof}$ $(\equiv N_{\rm data} - M)$, 
suitable for the least-square fitting procedure that we will
carry out for $M = 10$.

The pseudo data are generated by following the above recipe, 
as will be described in Subsection~\ref{sec-numpd}.

\subsection{Errors and Uncertainties}
\label{sec-estim}

In this subsection we discuss how to estimate the errors 
involved in the determination of density distributions. 
There are three kinds of errors;  
the completeness error, the systematic error, and the statistical error 
\cite{FN:NP}. 

\subsubsection{Completeness Error}
\label{sec-compl}

The completeness error is a measure of how closely the density
distribution is reconstructed by the use of a finite number of the basis
functions \cite{FN:NP}.   We introduce the relative completeness error as
\begin{eqnarray}
   \Delta \rho_{M}^{\rm com}(r)  
        \equiv | \rho_{\rm true}(r) - \tilde{\rho}_M(r) |
         \left/ \rho_{\rm true}(r) \right. ,
\label{eq:compl1}
\end{eqnarray}
where
\begin{equation} 
  \tilde{\rho}_M (r) = \sum_{n = 1}^{M} \tilde{C}_n \; f_n (r). 
\label{eq:expansion}
\end{equation}
$\{\tilde{C}_n\}$ is determined to minimize 
$(\rho_{\rm true}(r) - \tilde{\rho}_M(r))^2$ 
under the variations of $\tilde{C_n}$'s. 
Note that $f_n (r)$'s are linearly independent set, 
but are not necessarily orthogonal among each other. 
Since we use a finite number of the basis functions, 
the completeness of the set is not the issue here. 

$\{\tilde{C}_n\}$ is given by  
\begin{eqnarray}
  \tilde{C}_m = \sum_{n = 1}^M F_{m, n}^{-1} v_n,  
\label{eq:kg13p}
\end{eqnarray}
as the solution of 
\begin{eqnarray}
  \sum_{m = 1}^{M} F_{n, m} \tilde{C}_m = v_n, 
\label{eq:kg11p}
\end{eqnarray}
where
\begin{eqnarray}
 F_{n, m} &\equiv&  4 \pi \int_0^\infty r^l dr \; f_n (r) f_m (r), \\
 v_n &\equiv& 4 \pi \int_0^\infty r^l dr \; f_m (r) \; 
\rho_{\rm true}(r). 
\label{eq:kg12p}
\end{eqnarray}
Here, $l$ is the phase-volume parameter to specify the relationship 
among the basis functions, as described in Appendix~\ref{sec-kgauss}.  
In this work, we choose $l = 0$.

The relative completeness error introduced here will provide a useful 
means for assessing how well the surface distribution is determined.
In this work, we use the Kamimura-Gauss basis functions for $\{ f_n(r)\}$.

\subsubsection{Statistical Uncertainties and Systematic Errors}
\label{sec-statis}

When the true value, $X_{\rm true}$, is known, the mean square error of
a measured value, $X_{\rm data}$, is defined as \cite{Cowan},
\begin{eqnarray}
  \langle (X_{\rm data} - X_{\rm true})^2 \rangle 
   &=& \langle (X_{\rm data} - \langle X_{\rm data} \rangle )^2 \rangle 
   + ( \langle X_{\rm data} \rangle - X_{\rm true})^2,  \\
   &\equiv& \sigma^2 + b^2,
\label{eq:sta1}
\end{eqnarray}
where the bracket, $\langle A \rangle$, implies the sample mean of $A$.
$\sigma$ is the statistical error, and $b$ is the systematic error or the
bias. Generally, the error contained in data is an independent sum of the
statistical and systematic errors.
The statistical error shows how large the fitted results fluctuate 
around the sample mean, while the systematic error shows the deviation
of the sample means from the true value \cite{Cowan}. 

In our case, the statistical error of the fitted cross sections
at a $\theta_{\rm cm}$ is
\begin{eqnarray}
  \sigma_X^2(\theta_{\rm cm}) &\equiv& 
   \langle \left( {d\sigma^{\rm fit} \over d \Omega}
  - \langle {d\sigma^{\rm fit} \over d \Omega} \rangle 
            \right)^2 \rangle, 
\label{eq:sta2}
\end{eqnarray}
and the systematic error is
\begin{eqnarray}
  b_X^2(\theta_{\rm cm}) &\equiv&  
    \left( \langle {d\sigma^{\rm fit} \over d \Omega} \rangle
         - {d\sigma^{\rm true} \over d \Omega} \right)^2.   
\label{eq:sta3}
\end{eqnarray}
The sample mean denoted as $\langle \cdots \rangle$ in Eqs.~(\ref{eq:sta2}) 
and (\ref{eq:sta3}) at $\theta_{\rm cm}$ is explicitly 
\begin{eqnarray}
  \langle {d\sigma^{\rm fit} \over d \Omega} \rangle 
   \equiv {1 \over n} \sum_{i = 1}^n
          {d\sigma_{i}^{\rm fit} \over d \Omega}, 
\label{eq:sta4}
\end{eqnarray}
where the number of the pseudo-data set $n$ is 25. 

We define the statistical and systematic errors for the fitted density
distributions in the similar way.
\begin{eqnarray}
  \sigma_{\rho}(r)^2 &\equiv& \langle \left( \rho_M^{\rm fit}(r) 
                   - \langle \rho_M^{\rm fit}(r) \rangle \right)^2 \rangle, \\
  b_{\rho}(r)^2 &\equiv&  \left( \langle \rho_M^{\rm fit}(r) \rangle
                   - \rho_{\rm true}(r)  \right)^2, 
\label{eq:sta5}
\end{eqnarray}
where the sample mean is defined in the same way as in Eq.~(\ref{eq:sta4}).

\newpage
\setcounter{equation}{0}
\section{Numerical Calculations and Discussions}
\label{sec-nume}

\subsection{Completeness Error}
\label{sec-numcom}

We first examine the completeness error, 
Eq.~(\ref{eq:compl1}), of the Kamimura-Gauss basis functions,
taking $^{58}$Ni density as an example.  
The true density 
distribution, $\rho_{\rm true}(r)$, is taken to be of the three-parameter 
Fermi distribution form \cite{DeV:ATO},  
\begin{eqnarray}
  \rho(r) = \rho_0 \;{1 + w (r / c)^2 \over 1 + \exp \{(r - c)/a \}}, 
\label{eq:numcom1}
\end{eqnarray}
where $w = -0.1308$, $c = 4.3092$~[fm], and $a = 0.5169$~[fm].
$\rho_0$ is the normalization constant determined by  
\begin{eqnarray}
 \int d{\bf r} \; \rho({\bf r}) = 
 4 \pi \int_0^\infty r^2 dr \; \rho(r) = A.
\label{eq:numcom2}
\end{eqnarray}

Using Eq.~(\ref{eq:numcom1}) above as $\rho_{\rm true}(r)$ 
in Eqs.~(\ref{eq:kg13p}) - (\ref{eq:kg12p}), we can determine 
$\{\tilde{C_n}\}$, or $\tilde\rho_M(r)$ from Eq.~(\ref{eq:expansion}), 
and thus $\Delta \rho_{M}^{\rm com}(r)$ using Eq.~(\ref{eq:compl1}).
Our basis functions, the Kamimura-Gauss basis functions, however, depend 
on the size parameters, $r_1$ and $r_M$, defined in Eq.~(\ref{eq:form4}).  
All quantities involved for determining $\Delta \rho_{M}^{\rm com}(r)$, 
such as $\{\tilde{C_n}\}$, thus also depend on these parameters.  
We determine the values of $r_1$ and $r_M$ for a fixed $M$, so as to 
also minimize 
\begin{equation}
  \{ \rho_{\rm true}(r) - \tilde{\rho}_M(r) \}^2, 
\end{equation}
for the variations of $r_1$ and $r_M$.  This minimization is carried 
out numerically.

Table~1 lists the best-fit values of $r_1$ and $r_M$ for different $M$,
and Fig.~2a) shows the corresponding relative completeness error, 
$\Delta \rho_{M}^{\rm com}(r)$.  The radial range for the fit is 
$R = 10$~[fm]. 
In this completeness calculation, we impose no normalization condition on 
$\tilde{\rho}_M(r)$, and thus the integrated value of $\tilde{\rho}_M(r)$ also 
serves as a measure of goodness of $\tilde{\rho}_M(r)$.  The integrated value, 
$A_M$, is listed also in Table~1.  Note that the exact integrated value 
is 58 in this case. 

Judging from the $A_M$ value in Table~1 and 
$\Delta \rho_{M}^{\rm com}(r)$ in Fig.~2a), we see that the appropriate 
$M$ would be greater than 10.  As we will see in Subsecion~\ref{sec-numpd}, 
the uncertainty argument gives $M$ to be about 10.  We will choose $M$ $= 10$ 
in the rest of this work. 

Figure~2a) shows 
that $\Delta \rho_{M}^{\rm com}(r)$ increases in average 
as $r$ increases.  For $M$ $= 10$, $\Delta \rho_{M}^{\rm com}(r)$ increases 
to about 0.01 around 7~[fm] from 0.0001 at the central region.  Though the 
increase is appreciable, it would not be alarming as the density is down 
by the factor of about 1000 around 7~[fm], compared to the central region 
of the density. 
This behavior is quite similar to that of the Fourier-Bessel basis functions
(see Appendix~\ref{sec-fb}). 

We observe that while the $r_1$ value depends strongly on $M$, the $r_M$ 
rather weakly.  Because of this strong dependence, one may wonder how 
precisely the size parameters have to be chosen while keeping the 
completeness error to be reasonably small.  In order to address this 
issue, we choose a fixed set of the size parameters,  
$r_1$ $= 0.5$~[fm] and $r_M$ $= 6.0$~[fm], which deviate from the best-fit 
values for all $M$ shown in Table~1.  
Figure~2 b) illustrates 
$\Delta \rho_{M}^{\rm com}(r)$ for this choice of the size parameters. 
Though $\Delta \rho_{M}^{\rm com}(r)$'s in this case are larger than those  
for the best-fit parameter values, 
we consider $\Delta \rho_{M}^{\rm com}(r)$ in the order of 0.01 to be  
tolerable.  $A_M$ comes out to be quite close to the exact value, 
off by 3~[\%] even in the worst case of $M$ $= 5$.

The mild dependence of the completeness error on the size parameters 
will be further demonstrated as we apply the same best-fit size parameters 
for $^{58}$Ni to the case of $^{78}$Ni.  As described in 
Subsection~\ref{sec-appl}, we can exploit this mild dependence when we 
extend the model-independent analysis from the relatively well-known stable 
nuclei to poorly-known neutron-rich nuclei by keeping the same size 
parameters. 

\begin{table}
\begin{center}
\begin{tabular}{|c|c|c|c|}   
\hline
$M$ &  $r_1$~[fm]  &  $r_M$~[fm] & $A_M$ \\
\hline
 5  &   0.6        &   5.8       & 56.255   \\
10  &   1.0        &   5.5       & 58.030   \\
15  &   0.7        &   5.6       & 57.999   \\
20  &   0.3        &   5.7       & 58.000   \\
\hline
\end{tabular}
\end{center}
\label{tab:compl_kg}
\caption{Dependence of the best-fit parameters of $r_1$ and $r_M$ 
in the Kamimura-Gauss basis functions and of the integrated value of the density 
$A_M$ on the number of the basis functions $M$.   
The radial range of the fit is $R$ $= 10$~[fm]. 
}
\end{table}

The Kamimura-Gauss basis functions are not orthogonal among each other. 
Generally, however, the 
minimum point of the least-square does not change in the $M$ dimensional 
space of the basis functions under a rotation of the space coordinates 
attached to the basis functions.  The minimum point thus remains the same  
after the Kamimura-Gauss basis functions are orthogonalized 
by a linear transformation, 
say, through Schmidt orthogonalization method.  Note that this is the case 
even if a nonlinear transformation were to be used.  The error matrix 
at the minimum point does change, of course, under a transformation, linear 
or otherwise, by altering the correlated nature of the uncertainties.  
In the preceding discussions, we have addressed to this issue by performing 
numerical calculations by artificially moving away from the minimum point.  
This approach would be more reasonable, as we do not know 
the true distribution and thus the exact location of 
the minimum point in practical applications, neither. 

Before closing this subsection, 
we show the dependence of the cross sections on $M$ in Fig.~3. 
The Kamimura-Gauss basis functions give satisfactory convergence 
for $M$ $\ge 10$. 
To go beyond $20^{\circ}$ it gives nice diffraction patterns, 
especially for $M$ $\ge 15$. 
The solid curve ($M$ $= 20$) is on another dotted curve which is drawn 
by using the original $\rho_{\rm true}(r)$ for $\theta_{\rm cm}$ $< 20^{\circ}$, 
and slightly differs for $\theta_{\rm cm}$ $> 20^{\circ}$. 
This behavior is better than that of the Fourier-Bessel basis functions. 
(see Appendix~\ref{sec-fb}).

\subsection{Pseudo Data}
\label{sec-numpd}

As we described in Subsection~\ref{sec-pdata}, we generate 25 pseudo data 
at each scattering angle.  The values of the pseudo data change when the 
yield parameter $B_0$ is changed.  We carry out our analysis for three 
different values of $B_0$ $= 0.1$, 1.0, and 10~[mb/str].  Each value of 
$B_0$~[mb/str] provides different sets of the pseudo data, each set 
consisting of 25 data at each scattering angle.  We denote the set 
for $B_0$ $= 0.1$~[mb/str] to be the pseudo-data-set group A, 
the set for $B_0$ $= 1.0$~[mb/str] to be the group B, and   
the set for $B_0$ $= 10.0$~[mb/str] to be the group C. 
We take the minimum uncertainty $\epsilon_{\rm min}$ $=$ 2.0, 
$3.3$, and $6.7$~[\%] for pseudo-data-set group A, B and C, respectively 
(Table~2).
The definitions of these quantities are given in Appendix~\ref{sec-small}.

\begin{table}
\begin{center}
\begin{tabular}{|c|c|c|c|c|}   \hline
Group & $B_0$~[mb/str] & $\epsilon_{\rm min}$~[\%] & $\Delta \theta_{\rm min}$~[mrad] 
      & Rate~[sec$^{-1}$]  \\ 
\hline
 A    &    0.1          &      2.0          &    $2 \times 0.3$  
      &  $10^3$ \\
 B    &    1.0          &      3.3          &    $2 \times 0.3$  
      &  $10^2$ \\
 C    &   10.0          &      6.7          &    $2 \times 0.3$  
      &  $10$ \\
\hline
\end{tabular}
\end{center}
\caption{Parameters of the pseudo-data-set group A, B, and C, respectively. 
$B_0$ is the magnitude of cross section when 
the yield count is $N_0$. $N_0$ $= 10$ in this work. 
$\epsilon_{\rm min}$~[\%] is the minimum uncertainty.
$\Delta \theta_{\rm min}$~[mrad] is the minimum step. 
}
\end{table}

As an example, we plot a pseudo-data set of 
the pseudo-data-set group A in Fig.~4. 
The solid curve is the same as in Fig.~1. 
The kinematical conditions are:
\begin{eqnarray}
  \theta_{\rm max} \le 25^{\circ} ~~
 &\Leftrightarrow& ~~ q_{\rm max} \le 750 [{\rm MeV}].\\
 \Delta \theta \simeq 1.0^{\circ} &\Leftrightarrow& 
 \Delta q \simeq 20 [{\rm MeV}], 
\label{eq:numpd1b}
\end{eqnarray}
where $\Delta q$ is an increment in momentum transfer. 
Since we take a step in $\theta$ smaller than $1^{\circ}$ 
in the forward direction, $\Delta q$ can be smaller. 

Some of the parameters of the expansion of the density distribution 
are related to the kinematical conditions of the experiment 
\cite{FN:NP,Dre:NPA,BH:PRC,HB:NP}. 
The followings are the rough estimates for them: 
$r_{\rm max}$ is the maximum distance to probe by the experiment, and 
$r_{\rm max}$ $\simeq \pi/\Delta q$ $\simeq 30$~[fm]. 
We take $R$ $= 10$~[fm]~$(\le r_{\rm max})$. 
Since we are interested in the surface region, not in the tail, 
this value of $R$ is enough to cover the region. 
This choice is consistent with the charge density distribution studies
for the medium-heavy nuclei, $^{50, 52, 54}$Cr \cite{Lit:PRC}. 
The resolution of the fitted density distribution 
is  $\Delta r$ $= \pi/q_{\rm max}$ $\simeq 0.85$~[fm].
The dimension of the model space is $M$ $\simeq q_{\rm max} R / \pi$. 
We estimate $M$ $= 10 \sim 15$.

In order to make our discussions more realistic, we compare different 
pseudo-data-set groups with experimental setups expected at the RIBF. 
We first introduce two quantities, 
the counting rate, $J(\theta_{\alpha})$ [sec$^{-1}$], and  
the time for measurements, $T$~[sec]. 
The multiplication of the two gives the yield count,   
\begin{eqnarray}
  N_{\rm yield}(\theta_{\alpha}) = J(\theta_{\alpha}) \; T 
\label{eq:numpd1}
\end{eqnarray}
in Eq.~(\ref{eq:pd1}). 
$J(\theta_{\alpha})$ is expressed as 
\begin{eqnarray}
  J(\theta_{\alpha}) = N_{\rm tar} \; j_{\rm beam} \; 
  {d\sigma^{\rm true}_{\alpha} \over d\Omega} \Delta \Omega, 
\label{eq:numpd2}
\end{eqnarray}
where $N_{\rm tar}$ is the number of the particles in the reaction region 
of the target, 
$j_{\rm beam}$~[cm$^{-2}$ sec$^{-1}$] is the current of incident particles.  
$d\sigma^{\rm true}_{\alpha}$ $/ d\Omega$~[cm$^{2}$ str$^{-1}$] 
is the cross section at $\theta_{\alpha}$, 
and $\Delta \Omega$~[str] is the solid angle covered by the detector.

At the RIBF, experiments of proton-nucleus elastic scattering are performed  
in the inverse kinematics that the incident nuclear beam collides with 
protons in the target in the laboratory frame. 
Consider a CH$_2$ target with the thickness of 10~[mg cm$^{-2}$]. 
In terms of the Avogadro number $N_A$ $= 6.03 \times 10^{23}$~[mol$^{-1}$], 
the number of the protons in the target is  
$N_{\rm tar}$ 
$= 2 \times N_A$ $\times (10\times 10^{-3}/14)$
$= 2 \times 0.43 \times 10^{21}$~[cm$^{-2}$]. 
Let us assume the radius of the incident beam to be 1~[cm] and  
the number of the detectors to be 10 covering 50~[mstr] each (that is, 
$\Delta \Omega$ $= 0.5$~[str]), and also take  
$d\sigma^{\rm true}_{\alpha}$ $/ d\Omega$ $= 0.1$~[mb/str], 
and the rate of particle production to be $10^3$~[sec$^{-1}$] 
(that is, $j_{\rm beam}$ $= 10^3/\pi$~[cm$^{-2}$ sec$^{-1}$]). 
$J(\theta_{\alpha})$ is then in this case 
\begin{eqnarray}
  J(\theta_{\alpha}) &=& \pi~[{\rm cm}^2] 
      \times 2 \times 0.43 \times 10^{21}~[{\rm cm}^{-2}]\; 
     \times {10^3 \over \pi}~[{\rm cm}^{-2}{\rm sec}^{-1}] \nonumber \\
   &{}& ~~  \times 10^{-28}~[{\rm cm}^2{\rm str}^{-1}]
     \times 0.5~[{\rm str}]  \nonumber \\
  &=& 4.3 \times 10^{-5}~[{\rm sec}^{-1}]. 
\label{eq:numpd3}
\end{eqnarray}
If we suppose 1-week measurement, {\it i.e.}, $T$ $= 6.048 \times 10^5$~[sec], 
we have $N_{\rm yield}(\theta_{\alpha})$ $= 26.0$, or in the order of 
10 counts. 
Therefore, 0.1~[mb/str] roughly corresponds 
to a 10 yield for the above conditions, 
which is pseudo-data-set group A.
The correspondence of the rate of particle production 
to various pseudo-data-set groups is summarized in the Table~2. 

Note that the expected production rate of $^{70}$Ni at the RIBF is about 
$10^3$~[sec$^{-1}$], while that of $^{78}$Ni is about $10$~[sec$^{-1}$]. 
The former corresponds to the pseudo-data-set group A 
and the latter to the group C.
These correspondences are still valid if the target thickness becomes 
5~[mg cm$^{-2}$], because 
we have estimated them by considering $N_{\rm yield}(\theta_{\alpha})$ $= 26.0$ 
to be the order of 10 counts.

\subsection{Least-Square fitting}
\label{sec-fit}

For the three pseudo-data-set groups of proton-$^{58}$Ni elastic scattering 
generated in the 
previous subsection, we perform least-square fittings of Eq.~(\ref{eq:form3}) 
in Subsection~\ref{sec-form} to obtain the density distributions, $\rho(r)$. 
In the fittings, the normalization of $\rho(r)$ is treated as one of 
the pseudo data with an uncertainty of about 0.2~[\%], far smaller than the 
uncertainties assigned to the other pseudo data.  Note that the normalization 
of the basis functions is automatically taken care of in this way, 
because such a high precision datum works as a constraint. 
The numerical minimization is done by the use of MINUIT \cite{minuit}. 

There is another usage of the normalization. 
Without putting a high precision number to the normalization, 
one may make the fitting first, and calculate 
the normalization after the fitting is completed. 
This normalization should fluctuate around the true number, 
and the degree of this fluctuation could work as a measure of the 
goodness of the fitting. 
We do not take this way, because we are 
interested in the statistical and systematic errors of the density distribution 
for the fixed normalization.

The numerical results for 
the pseudo-data-set group A, B, and C are shown in Figs.~5, 6, and 7, 
respectively, for $R$ $= 10$~[fm] and $M$ $= 10$. 
As the particle production rate decreases from Fig.~5 to Fig.~7 (see Table~2), 
quality of the pseudo data becomes worse accordingly. 
Each figure consists of a)~25 fitted cross sections (shown with solid curves) 
with all pseudo data (crosses with bars), 
and b)~25 fitted density distributions, (solid curves), 
with the original density distribution (the dotted curve). 
The density distributions are shown for $r^2$ $\rho_M^{\rm fit}(r)$  
including the radial phase space of $r^2$.
It is this form that the data actually constrain.
In the figures we show 
the statistical and systematic errors divided by the true density distributions. 
One can see that the relative uncertainties of the fitted density
around the true density itself without $r^2$ from those figures. 
The average values of the least square divided 
by the number of degrees of freedom, $\chi^2/$DOF, 
and of the RMS radius, $\langle r^2 \rangle^{1/2}$, 
obtained from the 25 different pseudo-data sets are also shown in the figures. 
The number of DOF is $(\equiv N_{\rm data}$ $- M)$.

For all pseudo-data-set groups, 
the density distributions, $r^2$ $\rho_M^{\rm fit}(r)$, 
have larger uncertainties in the central region than in the surface region. 
This result is in agreement with the previous 
works \cite{PRC:Bri,Ray:PRC,Ray:PRC79,Hof:PRC,Str:PRC}, 
and is analogous to what was found in the determination of charge density 
distributions (for example, \cite{BM:NPA,Dre:NPA}). 
The original distribution drawn by the dotted curve is included in the error-band, 
which signifies that our procedure consistently provide probable 
density distributions within the uncertainty. 
Even with a pseudo-data-set group of lower yields, 
it seems that the surface has smaller uncertainty than the center. 
It is reasonable that the surface region of $\rho_M^{\rm fit}(r)$ 
is determined, 
because the cross section carries the information of 
this region \cite{ADL80}.  

In each case, $\chi^2/$DOF is about unity. 
For the pseudo-data-set group B
we have checked that the least square, $\chi^2$, of each pseudo data set 
distributes according to the $\chi^2$-distribution 
(Appendix~\ref{sec-distri}). 
This confirms that our numerical simulations make sense statistically. 
The $\langle r^2 \rangle^{1/2}$ of 
each pseudo-data-set group has less than 2 \% uncertainty,  
including the original value of 3.764~[fm]. 

Comparing Figs.~5, 6, and 7, we find that while the differential cross 
sections fluctuate for $\theta_{\rm cm} > 20^{\circ}$, the nuclear densities
fluctuate for $r < 2$[fm], and that the fluctuations become more 
appreciable again as the quality of the pseudo data decreases.
Note that when the number of terms of the Kamimura-Gauss basis functions, $M$, is 
increased, the $\chi^2/$DOF becomes closer to unity but the fluctuations 
increase, as similar observation was made previously 
\cite{FN:NP,BH:PRC,HB:NP}. 

In order to see explicitly the role of large $\theta_{\rm cm}$ or large-$q$ 
pseudo data, we remove the pseudo data 
of $\theta_{\rm cm}$ $> 15^{\circ}$ from the pseudo-data-set group C, 
and call the truncated data set to be the group C'. 
Figures 8 a) and b) show that the fluctuations by 
the pseudo-data-set group C' are greater than those by 
the group C both in the cross sections and in the density.  
Note that the mean value and the variance of $\langle r^2 \rangle^{1/2}$ 
changes much less, because the density in the surface region mainly affects  
$\langle r^2 \rangle^{1/2}$. 

The statistical uncertainty and the systematic error are examined in 
Figs.~9 and 10.  We see that as the quality of the pseudo data decreases, 
the statistical uncertainty tends to increase but the systematic error 
does less.  Furthermore, Fig.~9 shows that both of the uncertainty and 
the error in the surface region of our interest, $r=$ $3 \sim 5$~[fm], 
are smaller than in the other regions, 
which is in consistent with the previous works \cite{Ray:PRC79,Hof:PRC}. 
This is one of our major results.

The quantities examined 
here are the relative quantities.  One should not be alarmed with the 
somewhat large values for $r > 6$~[fm], because they are associated with 
quite small values of the density.  A similar caution is needed with the 
cusps appearing in Fig.~10, because they are associated with the dips in  
the cross sections.  The cusps are expected to reduced once the Coulomb 
interaction is included in the analysis.

\subsection{Neutron-rich nuclei}
\label{sec-appl}

We now examine an extension of our analysis to the case of neutron-rich 
nuclei by taking the case of the proton-$^{78}$Ni elastic scattering 
as an example.  Here, we use the pseudo-data-set group C 
to generate its pseudo data, 
as the expected particle production rate of $^{78}$Ni 
at the RIBF, $10$~[sec$^{-1}$], corresponds to the case of the group C. 
As the true density, we take the vector part of the relativistic 
mean-field density \cite{Men:PLB}.
Figure~11a) shows the true density distribution of $^{78}$Ni together 
with the distribution fitted by the use of fifteen ($M = 15$) 
Kamimura-Gauss basis 
functions.  They agree each other quite closely, as they could not be 
identified separately by eye.
$\langle r^2 \rangle^{1/2}$ $= 4.181$~[fm] is obtained from these density 
distributions. 

We use the same size parameters, $r_1$ $= 0.7$~[fm], and $r_M$ $= 5.6$~[fm], 
as those that we used for $^{58}$Ni with $M$ $= 15$ (Table~1). 
The completeness errors for  $M$ $= 10$ and 15 are shown in Fig.~11 b),  
which are very close to those for $^{58}$Ni. 
Therefore, once we determine the size parameters for a given $M$ 
for a known density distribution, such as that for stable nuclei, 
we could safely use them for the isotopes of the nucleus.

The numerical results of the least square fittings 
are shown in Fig.~12 a) and b). 
We use $R$ $= 10$~[fm] and $M$ $= 10$. 
The size parameters, $r_1$ $= 1.0$~[fm], and $r_M$ $= 5.5$~[fm], 
are the same values as those of $^{58}$Ni (Table~1). 
The pseudo data are fitted well, especially in the forward direction, 
as seen in Fig.~12 a). 
Figure~12 b) shows that the fitted density distributions fluctuate 
in the central region, but are stable in the surface region.
Unfortunately, $\langle r^2 \rangle^{1/2}$ obtained from the fittings 
is slightly smaller than that of the true distribution 
shown in Fig.~11 a), but within the fluctuation. 

Similarly to the discussions in the previous subsection, 
we also show the statistical uncertainty and the systematic error 
in Figs.~13 and 14.  Figure~13 confirms that the density distribution 
for this neutron-rich nucleus can be determined well in the surface region, 
which is consistent with the result obtained by using 
the Fourier-Bessel basis functions (Fig.~15).
$M$ $= 7$ is the best choice for this pseudo-data-set group
for the Fourier-Bessel basis functions of $R$ $= 10$~[fm].  

One should be careful for the statistical uncertainty in the central region. 
By comparing Fig.~13 with Fig.~15, 
one can see that the statistical uncertainty 
of the Kamimura-Gauss basis functions 
in the central region fluctuates much larger, 
about an order of magnitude, than the Fourier-Bessel ones.
As we mentioned in Introduction, 
since the statistical uncertainty should carry the information 
only on the quality of data, 
any basis function has to show the statistical uncertainty distribution 
of similar quality for the same data. 
The error estimation of the Kamimura-Gauss basis functions in the central region 
should be treated with caution. 

The behavior of these uncertainty and error is similar to that for the 
stable nucleus examined in the previous subsection. 
We thus expect that the surface region of nuclei could be better determined 
than the central region for relatively low intensity beam.

Before closing this section, we should add a comment. 
To discuss how the density distributions or the radii change in the isotope chain, 
the difference between the one from the neighboring nucleus is a good measure, 
because the model dependence could be canceled 
by the subtraction procedure \cite{AWT:NP81,Scl:PLB}. 
In this work we are interested in how precisely we can determine 
the density distribution 
itself of the neutron-rich nuclei with low intensity beam. 
We leave the problem concerning the discussion on such differences 
for our future work.

\newpage
\setcounter{equation}{0}
\section{Summary and Discussion}
\label{sec-summary}

The matter density distribution in the surface
region is shown to be determined well by a relatively low intensity beam 
in the order of $10$~[sec$^{-1}$], which is an expected intensity of 
the $^{78}$Ni beam at the RIBF of RIKEN.   

We have demonstrated how the determination of the density
distributions becomes poorer, especially in the central region,
as the quality of data becomes worse.  The demonstration is
made through a simulation by the use of three pseudo-data-set groups of  
proton-nucleus elastic scattering, corresponding to beam intensities
expected to be available at the RIBF.   Our discussion here is qualitatively 
in accord with what has been found in electron scatterings 
\cite{BM:NPA,Bat:ANP,Sic:LNP,Fro:MT,FN:NP,Dre:NPA}, 
and also with what has been done in proton scatterings \cite{Ray:PRC79,Hof:PRC}. 

The matter density distribution of neutron-rich nuclei is determined
by fixing the parameters in the basis functions to those of the neighboring
stable nucleus.  This procedure is expected to be a practical way of
treating neutron-rich nuclei, as the beam intensities of neutron-rich nuclei
tend to be less and the determination of their matter density 
distributions will be poorer.  
We have demonstrated this procedure to be feasible by taking
$^{78}$Ni as an example.

In this work, we have focused on the question, how well the nuclear density
distribution is determined from experimental data with the quality expected
at the upcoming facilities, {\it assuming} that the proton-nucleus
potential is known with no uncertainty.  The uncertainty for this must be
included in the practical evaluation of the uncertainty in the
determination of the matter density distribution.  The question of the potential
will require a close examination beyond the use of $t$-$\rho$ form,
especially as the incident energy becomes lower.  Also, a special attention
may have to be given to neutron-rich nuclei \cite{Var:pre}. 

We expect that the uncertainty
associated with the use of the eikonal form instead of the partial wave
decomposition is small, though the absolute magnitude itself could be
substantial.  
In order to clarify this and the uncertainty associated with the potential, 
they will be the issues that we wish to address in future work.

\begin{center}
 {\bf Acknowledgements}
\end{center}

We acknowledge I.Tanihata for his invaluable suggestions and comments,
and M. Kamimura and K.Yazaki for stimulating discussions. 
We thank T.Ohnishi, and T.Suda, and K.Katori for their assistance
on our generation of the pseudo data to be realistic to the RIBF facilities.
G.D.Alkhazov and A.Lobodenko informed us of their work \cite{Lom:NPA}, and
M.Oka provided us a computer code for orthogonalization of a function set. 
This work is supported under the Special Postdoctoral Research Program
at RIKEN and by the U.S. DOE at CSUN (DE-FG03-87ER40347) and
the U.S. NSF at Caltech (PHY00-71856 and PHY97-22428).

\newpage
\appendix
\begin{center}
  {\bf APPENDIX}
\end{center}
\setcounter{equation}{0}
\section{Kamimura-Gauss Basis functions}
\label{sec-kgauss}

Here we summarize the formulae associated with  
the Kamimura-Gauss basis functions, 
which have been used by Kamimura \cite{Kam:PRA} 
in various applications, for example, in few-body calculations \cite{Hi:NP}. 

The basis functions are expressed as 
\begin{eqnarray}
 f_n(r) = \exp ( - r^2 / r_n^2) \; \theta(R - r), ~~(n = 1, \cdots, M),
\label{eq:kg1}
\end{eqnarray}
where the size parameter, $r_n$, is determined through a geometrical 
progression, $r_n = r_1 a^{n - 1}$ $(n = 1, \cdots, M)$. 
$\theta(x)$ is the step function, and 
$R$ is a scale parameter constrained by the kinematical condition \cite{FN:NP}.
Note that the standard version of the Kamimura-Gauss basis functions has no 
restriction of $r_M < R$ \cite{Kam:PRA}. 

The inner product of two basis functions is defined by 
\begin{eqnarray}
  (f_n, f_{n + m})_l 
  &\equiv& 4 \pi \int_{0}^{\infty} r^l dr\; f_n(r) f_{n + m}(r) \\
  &=& {2 \pi \over \alpha_{n, m}^{(l + 1)/2}} 
      \{ \Gamma \left((l + 1)/2 \right) 
       - \Gamma \left((l + 1)/2, \alpha_{n, m} R^2 \right) \},
\label{eq:kg2}
\end{eqnarray}
where 
\begin{eqnarray}
  \alpha_{n, m} \equiv {1 \over r_n^2} + {1 \over r_{n + m}^2} 
                     = {1 \over r_n^2} \left( 1 + {1 \over a^{2m}} \right),
\label{eq:kg3}
\end{eqnarray}
and $l$ is the power of the phase volume and is set to be a non-negative 
integer.  $\Gamma(z)$ $(= \Gamma(z, 0) )$ is the Gamma function, and 
$\Gamma(z, p)$ is the incomplete Gamma function of the second kind. 
 
When the normalization is suitably chosen, the expression of the inner 
product becomes simple.  If $(\tilde{f_n}, \tilde{f_n})_{l = 2}$ $= 1$  
for $\tilde{f_n}(r)$ $= N_n \; f_n(r)$, we obtain
\begin{eqnarray}
 {1 \over N_{n}^2} = {\pi \over \sqrt{2}} \; r_m^3 \;
         \left( \Gamma(3/2) - \Gamma(3/2, 2 R^2/r_n^2) \right).
\label{eq:kg6}
\end{eqnarray}
Note that when $R \rightarrow \infty$, the expressions become greatly 
simplified:
\begin{eqnarray}
 {1 \over N_{n}^2} = {\pi^{3/2} \over 2 \sqrt{2}} \; r_n^3, 
\label{eq:kg7}
\end{eqnarray}
and 
\begin{eqnarray}
  (\tilde{f_n}, \widetilde{f_{n + m}})_{l = 2} 
  &=& {2 \sqrt{2} a^{3m/2} \over (1 + a^{2m})^{3/2}}. 
\label{eq:kg8}
\end{eqnarray}

The basis functions have the following properties: 
1) They are not orthogonal, but are linearly independent. 
One can easily verify the linear independence by demonstrating the Wronskian, 
to be non-zero except at the infinity. 
2) The derivative with respect to $r$ vanishes at $r$ $= 0$, and thus  
the derivative of the density expanded in terms of the basis functions  
is also zero at $r$ $= 0$. Note that this property is the same as that 
of the Gaussian basis functions used by Sick \cite{Sic:NP}.  
3) The coefficients of the Kamimura-Gauss basis functions can be negative, 
while Sick restricted the coefficients of his Gaussian basis functions 
to be positive \cite{Sic:NP}.  Actually, the coefficients in our case tend 
to appear almost equally in both signs. 
4) The set of the Kamimura-Gauss basis functions 
does not form a complete set, even if 
the infinite number of terms are used.  Note that a finite number of 
terms is always used in practice, and the issue of completeness 
is an academic question. 
5) They depend on two parameters, such as $r_1$ and $r_M$.

In our application, we choose $r_1$ and $r_M$ by minimizing  
$D(\tilde{C_n})$ of Eq.~(\ref{eq:kg9}) below. 
Though $r_1$ depends strongly on $M$, it is found 
to vary little for different isotopes. 
The expansion coefficients for the normalized basis functions, $\tilde{C_n}$, 
are determined by minimizing 
\begin{eqnarray}
  D(\tilde{C_n}) &\equiv& |\rho_{\rm true}(r) - \rho_M^{\rm fit}(r)|^2 \\
 &=&  |\rho_{\rm true}(r) - \sum_{n = 1}^{M} \tilde{C_n} \; \tilde{f_n}(r)|^2, 
\label{eq:kg9}
\end{eqnarray}
that is, by satisfying  
$\partial D(\tilde{C_m}) / \partial \tilde{C_m}$ $= 0$. 
This condition leads to 
\begin{eqnarray}
  \tilde{f_m}(r) \; \rho_{\rm true}(r) 
  = \left( \sum_{n = 1}^{M} \tilde{C_n} \; \tilde{f_n}(r) \right) 
    \tilde{f_m}(r), ~~(n, m = 1, \cdots, M).
\label{eq:kg10}
\end{eqnarray}
We thus obtain
\begin{eqnarray}
  \sum_{m = 1}^{M} F_{n, m} \tilde{C_m} = v_n, 
\label{eq:kg11}
\end{eqnarray}
where
\begin{eqnarray}
 F_{n, m} &\equiv&  4 \pi \int_0^\infty r^l dr \; \tilde{f_n}(r) \tilde{f_m}(r). \\
 v_n &\equiv& 4 \pi \int_0^\infty r^l dr \; \tilde{f_m}(r) \; \rho_{\rm true}(r). 
\label{eq:kg12}
\end{eqnarray}
Note that one can also follow the same procedure in the case of the 
coefficients, $\tilde{C_n}$, for non-normalized basis functions. 
The phase-volume parameter, $l$, is chosen by depending on how much weight 
one wishes to impose on the out-region.  In the present analysis, 
we choose $l$ $= 0$ in order to include the information 
of the whole region equally. 

Finally, Eq.~(\ref{eq:kg11}) is inverted numerically to yield
\begin{eqnarray}
  \tilde{C_m} = \sum_{n = 1}^M F_{m, n}^{-1} v_n. 
\label{eq:kg13}
\end{eqnarray}

\newpage
\setcounter{equation}{0}
\section{Eikonal Expressions}
\label{sec-xsec}

For the purpose of completeness and clarification, 
we summarize the eikonal expressions \cite{Glau:Lec}, 
which we have used. 

The differential cross section of the proton-nucleus elastic scattering
in the center-of-mass frame is expressed as 
\begin{eqnarray}
  {d\sigma_{\rm el} \over d\Omega}
  =  F_{\rm kin}\; |T({\bf q})|^2 
\label{eq:xsec1}
\end{eqnarray}
with the kinematical factor $F_{\rm kin}$ $= |{\bf p}|^2 / (2 \pi)^2$. 
${\bf q}$ $(= {\bf p} - {\bf p}')$ is the three-momentum transfer, 
where ${\bf p}$ and ${\bf p}'$ are the momenta of the incident 
and of the outgoing proton, respectively. 
$T({\bf q})$ is the $t$-matrix expressed in the eikonal form,
\begin{eqnarray}
 T({\bf q}) 
  &=& \int d{\bf b} \; \exp \{i {\bf q} \cdot {\bf b} \} \; 
      \left( 1 - \exp \{i \chi({\bf b})\} \right), \\
  &=& 2 \pi \int_{0}^{\infty} b db \; J_{0}(qb) \; 
      \left( 1 - \exp \{i \chi(b) \} \right), 
\label{eq:xsec2}
\end{eqnarray}
where $J_0(qb)$ is the zero-th order Bessel function.  
The profile function, $\chi({\bf b})$, is given by
\begin{eqnarray}
  \chi({\bf b})  
  &=& {2 \pi \over |{\bf p}|} \; f({\bf 0}) \;
          \int_{-\infty}^{\infty} dz' \; \rho({\bf b}, z'), 
\label{eq:xsec4}
\end{eqnarray}
where $\rho({\bf r})$ is the one-body matter density distribution, 
and ${\bf r}$ $= ({\bf b}, z)$. 
${\bf b}$ is the impact parameter vector, ${\bf b}$ $\perp {\bf p}$, and 
the integration in $z$ is taken in the direction 
of the incident momentum, ${\bf p}$. 
$f({\bf 0})$ is the nucleon-nucleon scattering amplitude 
in the forward direction, ${\bf q}$ $= {\bf 0}$. 
\begin{eqnarray}
  f({\bf 0})  
  &=& {|{\bf p}| \over 4 \pi} \; (i + \alpha) \; \sigma_{\rm NN}^{total}.
\label{eq:xsec5}
\end{eqnarray}
The expression of $\chi({\bf b})$ above corresponds to
the first-order optical potential in the Glauber approximation.
The Coulomb interaction is not included here, but could be included 
straightforwardly \cite{Glau:Lec,GM:NPB}.   

For the numerical calculation of Fig.~1, 
we use the parameters of Eq.~(\ref{eq:xsec5}) as follows \cite{LB}:
\begin{eqnarray}
 |{\bf p}_{lab}| &=& \sqrt{(K + m)^2 - m^2} = 1.75~[{\rm GeV/c}]. \\
  \alpha &=& -0.0687, ~~~
  \sigma_{\rm NN}^{total} = 43.1~[{\rm mb}], 
\label{eq:xsec6}
\end{eqnarray}
where $m$ is the nucleon mass, and $K$ is 
the kinetic energy of the incident proton, which is $K =$ 1.047~[GeV]. 
We use these parameters for all the numerical calculations in this work. 

We calculate the cross section of proton-$^{58}$Ni elastic scattering 
as an example in Subsection~\ref{sec-form} and Section~\ref{sec-nume}. 
We take the three-parameter Fermi distribution, Eq.~(\ref{eq:numcom1}), 
for the matter density distribution of $^{58}$Ni 
to create the pseudo data in Subsection~\ref{sec-numpd}, 
because the experimental data of differential cross sections 
of electron-nucleus elastic scattering for various nuclei are fitted well 
with this form, including $^{58}$Ni \cite{DeV:ATO}, 
and because the matter density distribution of $^{58}$Ni is considered to be 
quite similar to the charge density distribution. 
Actually, the experimental data of differential cross section 
of proton-$^{58}$Ni elastic scattering are fitted well 
with this form \cite{Alk:PLB}. 

One should be careful for large-$r$ behavior 
of the three-parameter Fermi distribution, Eq.~(\ref{eq:numcom1}), 
because it becomes negative for $r > r_c$ if $w$ is negative. 
That is, $\rho(r_c)$ $= 0$ $\Leftrightarrow$ $1 + w (r_c / c)^2$ $= 0$, 
for $r_c$ $= c / \sqrt{-w}$. 
For our choice of the parameters, $r_c$ $= 11.915$~[fm].
We put $\rho(r) = 0$ for $r > r_c$. 

With the fitted density distribution, 
$\rho_M^{\rm fit}(r)$, of Eq.~(\ref{eq:form1}) in terms of 
the Kamimura-Gauss basis functions defined by Eq.~(\ref{eq:form4}), 
we can perform the integration of the profile function in Eq.~(\ref{eq:xsec4}) 
analytically due to the nature of the Gaussian functions: 
\begin{eqnarray}
  \chi({\bf b})  
  &=& {2 \pi \over |{\bf p}|} \; f({\bf 0}) 
       \sum_{n = 1}^{M} C_n \; r_n \; \exp \{- {\bf b}^2/r_n^2 \}
\nonumber \\
  &{}& ~~~~~ \times
        \{ \Gamma (1/2) - \Gamma \left(1/2, (z_{\rm max}/r_n)^2 \right) \}, 
\label{eq:xsec4b}
\end{eqnarray}
where $\Gamma (1/2)$ $= \sqrt{\pi}$, and $z_{\rm max}^2$ $= R^2 - {\bf b}^2$. 
$\Gamma(z, p)$ is the incomplete Gamma function of the second kind. 
This integration reduces our numerical task for the cross sections, 
and brings us the numerically stable results.

\newpage
\setcounter{equation}{0}
\section{Fourier-Bessel Basis Functions}
\label{sec-fb}

Here we summarize some formulae concerning 
the Fourier-Bessel basis functions, 
and show its completeness error introduced in Subsection~\ref{sec-compl}. 
The Fourier-Bessel basis functions are defined by \cite{FN:NP}
\begin{eqnarray}
 f_n(r) = j_0(n \pi r / R) \; \theta(R - r),
\label{eq:fb1}
\end{eqnarray}
where $j_0(x)$ is the 0-th order spherical Bessel function, 
which is expressed by $j_0(x)$ $= \sin x / x$.
$R$ is a scale parameter determined by the discussion
under eq.~(\ref{eq:numpd1b}). 
$\theta(x)$ is the step function. 

If we define the inner product as 
\begin{eqnarray}
  (\tilde{f_n}, \tilde{f_m})_{l = 2}
  &\equiv& 4 \pi \int_{0}^{\infty} r^2 dr\; \tilde{f_n}(r) \tilde{f_m}(r), 
\label{eq:fb3}
\end{eqnarray}
we obtain the normalized version of this basis function as, 
\begin{eqnarray}
 \tilde{f_n}(r) = N_n \; f_n(r),  
\label{eq:fb2}
\end{eqnarray}
where $N_n$ $= \sqrt{n^2 \pi /(2 R^3)}$.

Since the Fourier-Bessel basis functions are orthogonal to each other, 
{\it i.e.}, $(\tilde{f_n}, \tilde{f_m})$ $= \delta_{n, m}$. 
we can easily obtain $\tilde{C_n}$ by the following equation: 
\begin{eqnarray}
   \tilde{C_n} = 4 \pi \int_0^\infty r^2 dr \; 
       \tilde{f_n}(r) \; \rho_{\rm true}(r).
\label{eq:fb4}
\end{eqnarray}

As in Subsection~\ref{sec-numcom}, 
we show the numerical results of $\Delta \rho_{M}^{\rm com}(r)$,  
defined by Eq.~(\ref{eq:compl1}), 
for the Fourier-Bessel basis functions of $R$ $= 10$~[fm] 
with the number of terms, $M$, being 5, 10, 15, and 20, in Fig.~16 a). 
This figure corresponds to Fig.~2 a). 
Here we also systematically determine the expansion coefficients 
of the Fourier-Bessel basis functions, $\tilde{C_n}$, of Eq.~(\ref{eq:fb4}). 
Roughly speaking, the magnitude of the coefficients becomes smaller 
as $n$ increases, because the overlap of $\tilde{f_n}(r)$ 
with $\rho_{\rm true}(r)$ becomes smaller due to the oscillatory nature 
of the basis functions as one can see from Eq.~(\ref{eq:fb4}). 

Although each term of the Fourier-Bessel basis functions 
and the Kamimura-Gauss ones shows different behavior in $r$,  
it is interesting to see that $\Delta \rho_{M}^{\rm com}(r)$ of the both bases 
show similar oscillating patterns in $r$. 
The convergence at each point becomes better as $M$ increases 
for the both cases, 
and the convergence for the Fourier-Bessel basis functions become better
beyond $M$ $= 15$ in contrast to the Kamimura-Gauss ones. 
Up to $M$ $= 15$, the both bases have almost the same quality of convergence. 
The uncertainties for $M$ $\ge 10$ are the order of $10^{-2}$ or less, 
and for $M$ $\ge 15$ are the order of $10^{-3}$ or less 
in the region of 1~[fm] $\le r$ $\le 5$~[fm],  
which covers the surface region of nuclei. 
The uncertainty becomes worse in the region of $r$ $> 5$~[fm], 
but it is still the order of $10^{-2}$. 

\begin{table}
\begin{center}
\begin{tabular}{|c|c|}   \hline
\multicolumn{2}{|c|}{$R = 10$ ~[fm]} \\
\hline
$M$ &  $A_M$    \\
\hline
 5  &  58.115  \\
10  &  57.974  \\
15  &  58.001  \\
20  &  57.999  \\
\hline
\end{tabular}
\end{center}
\label{tab:compl_fb}
\caption{Dependence of the integrated value of the density 
$A_M$ of the Fourier-Bessel basis functions on the number 
of the basis functions $M$.   
The radial range of the fit is $R$ $= 10$~[fm]. 
}
\end{table}

As in the completeness calculation in Subsection~\ref{sec-numcom}, 
we impose no normalization condition (\ref{eq:form2}) on 
$\tilde{\rho}_M(r)$, Eq.~(\ref{eq:expansion}), here, 
and thus the integrated value of $\tilde{\rho}_M(r)$ also 
serves as a measure of goodness of $\tilde{\rho}_M(r)$.  
As in Table~1 in Subsection~\ref{sec-numcom}, 
Table~3 lists the integrated value, $A_M$. 
Note that the exact integrated value is 58 in this case, too. 
As one can see from this table, 
$A_M$ stays in about 0.2~\%-fluctuation for all $M$. 
They are all tolerable. 
For $M$ larger than 5, $A_M$ can be determined within the error of 0.05~\%. 

Finally, 
we show the numerical results of the corresponding cross sections 
for each $M$ in Fig.~16 b). 
This figure corresponds to Fig.~3. 
The Fourier-Bessel basis functions give satisfactory convergence 
for $M$ $\ge 10$,  
similar to the Kamimura-Gauss ones, and, 
if we restrict ourselves to see $\theta$ $< 10^{\circ}$, 
we already obtain good convergence even for $M$ $= 5$. 
The solid curve ($M$ $= 20$) is almost on the dotted curve 
(drawn by the original $\rho_{\rm true}(r)$)  
for $\theta_{\rm cm}$ $< 20^{\circ}$, 
but unfortunately the diffraction pattern is broken 
for $\theta_{\rm cm}$ $> 20^{\circ}$. 
The other curves, $M$ $< 20$, are also differs from the dotted curve 
in this large-angle region. 

Such broken patterns of the Fourier-Bessel basis functions  
could be numerical artifacts coming from the numerical integration,  
which is almost inevitable for numerical fittings, 
while they do not appear for the Kamimura-Gauss basis functions. 
Thus, when we apply the Fourier-Bessel basis functions to those fittings, 
we should be careful for cross sections of the large scattering angle. 
They most probably affect the central region of the density distributions.  
For $\theta$ $< 20^{\circ}$ and $M$ $\ge 10$, 
both the Fourier-Bessel and Kamimura-Gauss basis functions 
give good convergence.

\newpage
\setcounter{equation}{0}
\section{Small Scattering Angles}
\label{sec-small}

Here we explain how to reduce the separation of pseudo data 
in $\theta_{\rm \alpha}$ from $1.0^{\circ}$ near the forward direction. 
The prescription to create pseudo data is written 
in Subsection~\ref{sec-pdata}. 
To make the way to shrink the step of pseudo data systematic,
we define the minimum uncertainty, $\epsilon_{\rm min}$.  
Since we are to assign a statistical uncertainty more than this, 
we restrict $\epsilon_{\rm min}^2$ $\leq 1/N_{\rm PD}(\theta_{\alpha})$  
at each data point, where 
$N_{\rm PD}(\theta_{\alpha})$ is the yield count of the pseudo data 
at $\theta_{\alpha}$, defined in the same way as in Eq.~(\ref{eq:pd2}). 
We obtain $\Delta \theta_{\alpha}$ by keeping 
\begin{eqnarray}
  \Delta \theta_0 \; {1 \over \epsilon_{\rm min}^2 }
  \simeq \Delta \theta_{\alpha} \; N_{\rm PD}(\theta_{\alpha}).  
\label{eq:pd4}
\end{eqnarray}
In the region where the theoretical cross section changes drastically 
this formula overestimates or underestimates the step. 

For large $N_{\rm PD}(\theta_{\alpha})$, 
$\Delta \theta_{\alpha}$ becomes smaller than a bin of the detector. 
In order to avoid such an unrealistic estimate, 
we define the minimum step, $\Delta \theta_{\rm min}$, 
and make $\Delta \theta_{\alpha}$ not less than that. 
The minimum step is expected to be 0.3~[mrad] 
$(= 17.2^{\circ}$ $\times 10^{-3})$ in the proton-fixed frame at the RIBF. 
Since our cross sections are calculated in the center-of-mass frame, 
we are careful for the inverse kinematics of the RIBF. 
We take $\Delta \theta_{\rm min}$ $= 2 \times 0.3$~[mrad], 
which is safe if we transform $\Delta \theta_{\rm min}$ 
in the proton-fixed frame to the center-of-mass frame. 
The requirement to identify a proton-nucleus elastic scattering 
makes the minimum step very small. 
In the forward direction $\theta$ can be determined precisely at the RIBF. 
Although the data have some uncertainty in $\theta$ 
in the large $\theta_{\rm cm}$ region, 
we neglect it in this work.

\newpage
\setcounter{equation}{0}
\section{Distribution of the Fitted Results}
\label{sec-distri}

According to the statistics, 
the least square, $\chi^2$, of Eq.~(\ref{eq:form3}) is known to obey 
the $\chi^2$-distribution defined by
\begin{eqnarray}
  f_\nu(\chi^2) = \left\{ \begin{array}{ll}
                  2^{-\nu/2}  \Gamma(\nu/2)^{-1} \;
                    (\chi^2)^{\nu/2 - 1} \exp (-\chi^2/2), 
                  &~~~ (\chi^2 > 0), 
\nonumber\\
                  0,    &~~~ (\chi^2 < 0), 
                 \end{array}
                     \right. 
\label{eq:numfit1}
\end{eqnarray}
where $\nu$ $(> 0)$ is the degree of freedom, 
provided that each term consisting the least square of Eq.~(\ref{eq:form3}), 
$(d\sigma_{\alpha}^{\rm data}/d \Omega$
$- d\sigma_{\alpha}^{\rm true}/d \Omega)$ 
$/\epsilon_{\alpha}$, obeys the standard normal distribution \cite{Cowan}. 

We use this fact to examine the consistency of our simulation 
by calculating the following quantity. 
\begin{eqnarray}
  F_\nu(\chi^2) = \int_0^{\chi^2} dv \; f_\nu(v). 
\label{eq:numfit2}
\end{eqnarray}
We compare it with our numerical results in Table~4. 

We take the pseudo-data-set group B as an example. 
The number of data points of each pseudo-data set
is 40, 41, and 42. 
Such a fluctuation of the number of data point occurs, 
because some pseudo data are rejected from the pseudo-data set 
when they become less than $10^{-2}$~[mb/str]. 
We use the pseudo-data sets of 41 data points, 
because this is the largest portion of the group. 
The degree of freedom becomes 31, as $M$ $= 10$. 
The case contains 15 pseudo-data sets out of 25. 

Table~4 shows that 
our fitted results distribute consistently with the calculated 
$\chi^2$-distribution, and that our simulations are in accord with 
statistical considerations.   

\begin{table}[b]
\begin{center}
\begin{tabular}{|c|c|c|}   \hline
$\chi_a^2$ & $F(\chi_a^2)$ calc. & $F(\chi_a^2)$ PDSG-B \\
\hline
   20      &    0.064         &      0.067 $(= 1/15)$   \\
   30      &    0.48          &      0.33  $(= 5/15)$   \\  
   40      &    0.87          &      0.87  $(= 13/15)$   \\
   50      &    0.98          &      1.00  $(= 15/15)$   \\
\hline
\end{tabular}
\end{center}
\caption{Comparison of calculated values of integrated $\chi^2$ distribution 
with the corresponding quantity of the pseudo-data-set group B.
The degree of freedom is 31. 
The bracket in the right most column shows the fraction 
which satisfies the condition. 
}
\end{table}

\newpage
\begin{figure}[h!]
  \begin{center}
    \includegraphics[width=40pc, keepaspectratio]{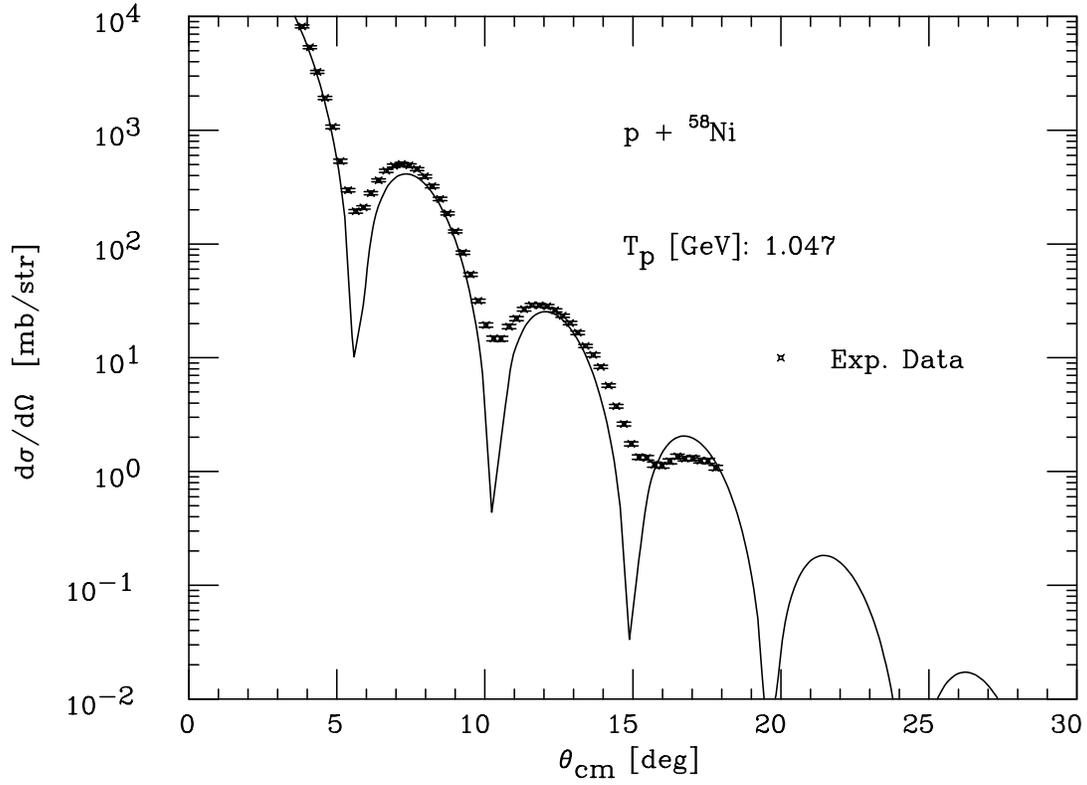}
  \end{center}
\caption{
Comparison of the theoretical cross section 
with the real experimental data. 
The solid curve is the theoretical cross section 
of the proton-$^{58}$Ni elastic scattering without the Coulomb interaction. 
The crosses with bar are the experimental data of SATURN at Saclay. 
The kinetic energy of the incident proton is 1.047~[GeV].
} 
\end{figure}
\begin{figure}[h!]
  \begin{center}
    \includegraphics[width=30pc, keepaspectratio]{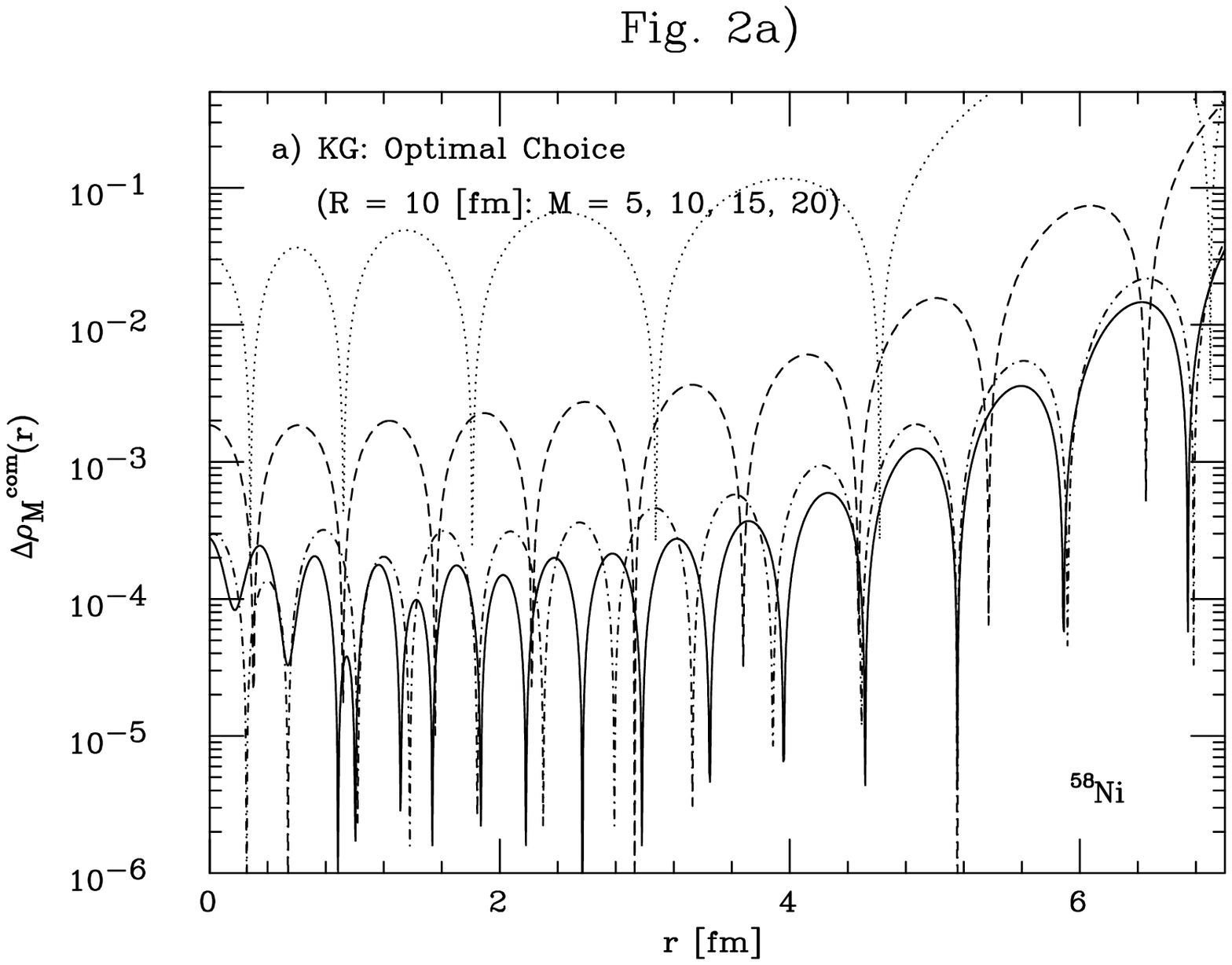}\\
    \includegraphics[width=30pc, keepaspectratio]{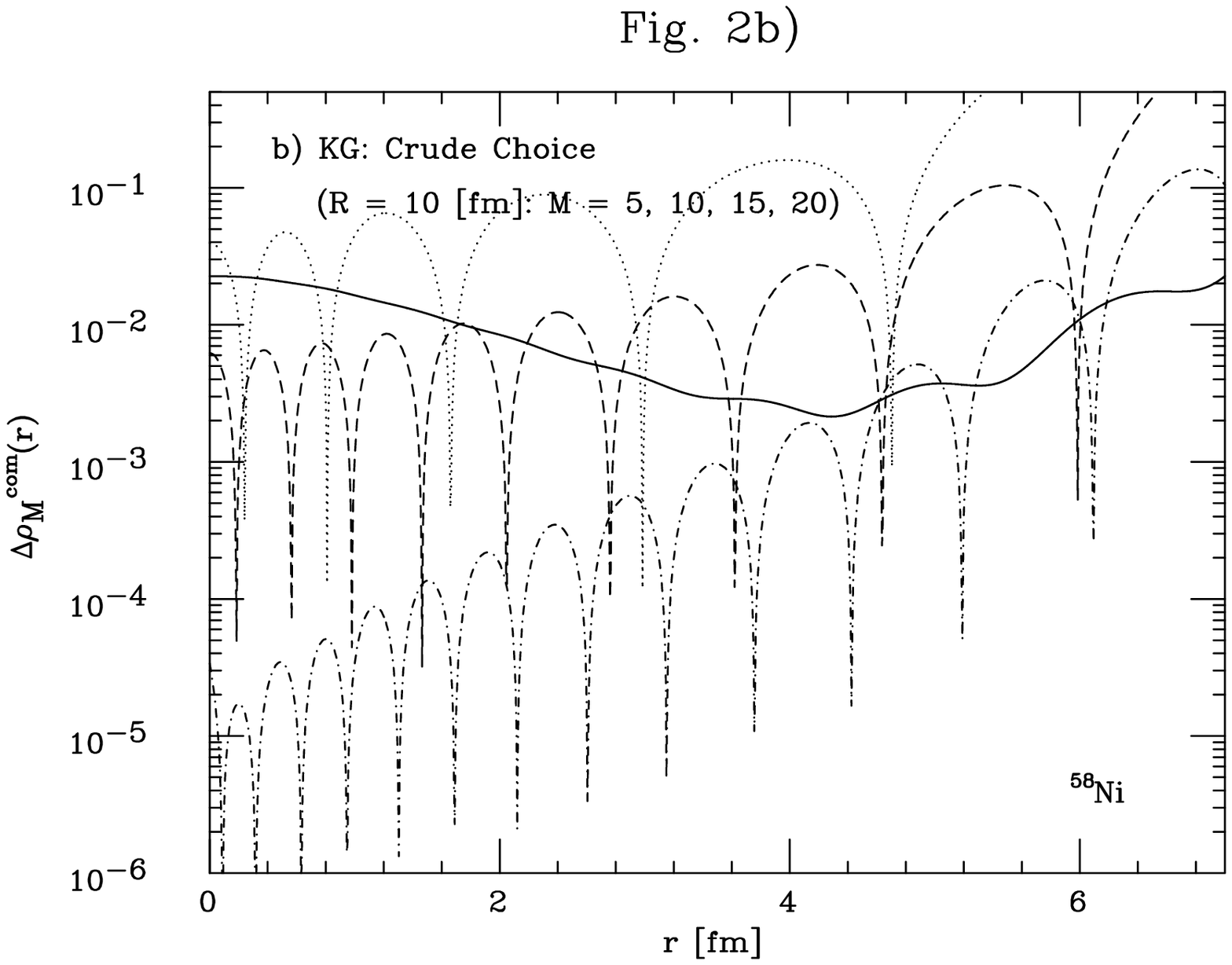}
  \end{center}
\caption{
$M$-dependence of the completeness error 
of the Kamimura-Gauss (KG) basis functions a)~with the best-fit parameters, 
b)~with a crude choice of $r_1$ and $r_M$, 
for $M =$ 5, 10, 15, and 20. $R$ $= 10$~[fm]. 
The dotted curve is for $M = 5$, 
the dashed one is for $M = 10$,
the dash-dotted one is for $M = 15$,
and the solid one is for $M = 20$.
} 
\end{figure}
\begin{figure}[h!]
  \begin{center}
    \includegraphics[width=40pc, keepaspectratio]{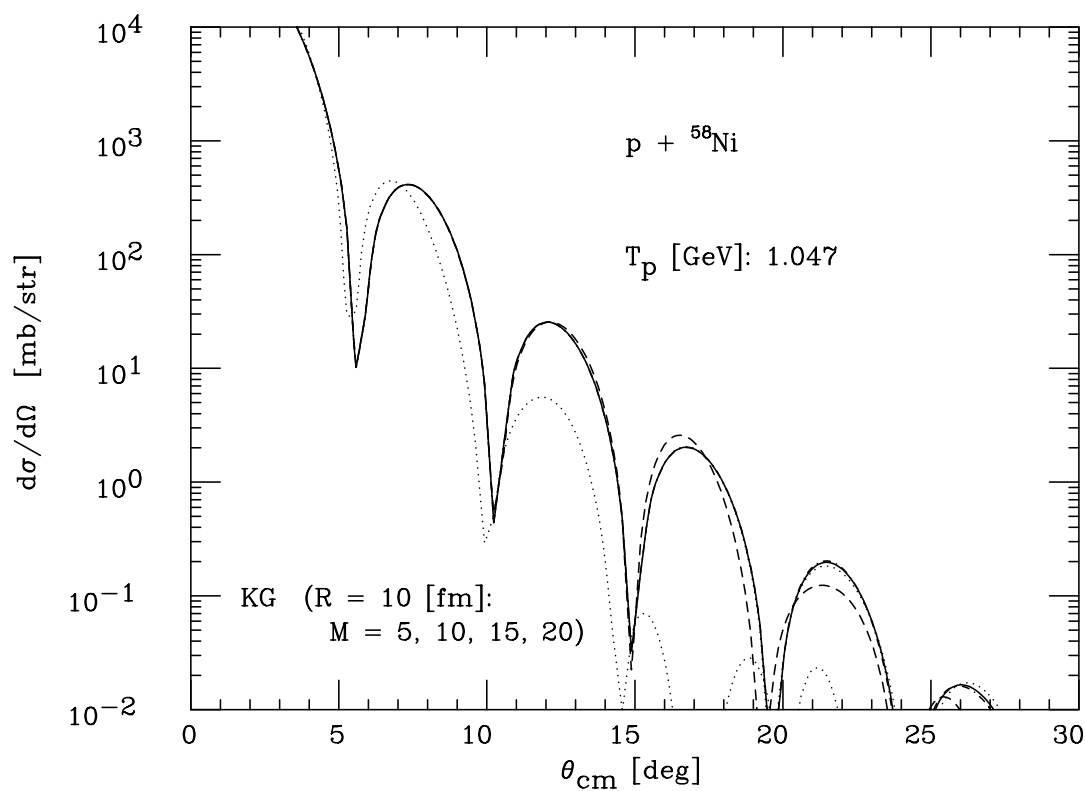}
  \end{center}
\caption{
$M$-dependence of the cross section 
with the Kamimura-Gauss (KG) basis functions 
for $M =$ 5, 10, 15, and 20. $R$ $= 10$~[fm]. 
The dotted curve is for $M = 5$, 
the dashed one is for $M = 10$,
the dash-dotted one is for $M = 15$,
and the solid one is for $M = 20$.
Another dotted curve, as it is quite difficult to be seen, 
is obtained by using the original $\rho_{\rm true}(r)$. 
} 
\end{figure}
\begin{figure}[h!]
  \begin{center}
    \includegraphics[width=40pc, keepaspectratio]{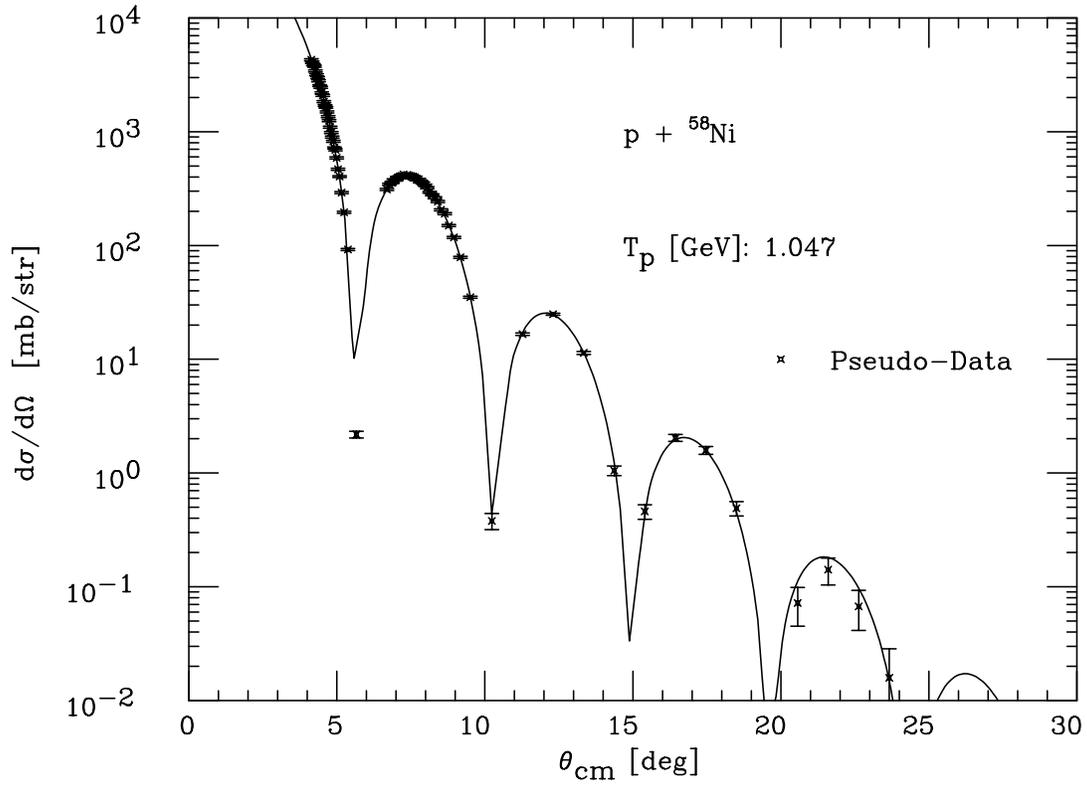}
  \end{center}
\caption{
An example of pseudo-data set of 
the pseudo-data-set group (PDSG) A. 
Those simulate a series of measurements of the cross section 
of proton-$^{58}$Ni elastic scattering. 
The solid curve is obtained by using the original $\rho_{\rm true}(r)$. 
} 
\end{figure}
\begin{figure}[h!]
  \begin{center}
    \includegraphics[width=30pc, keepaspectratio]{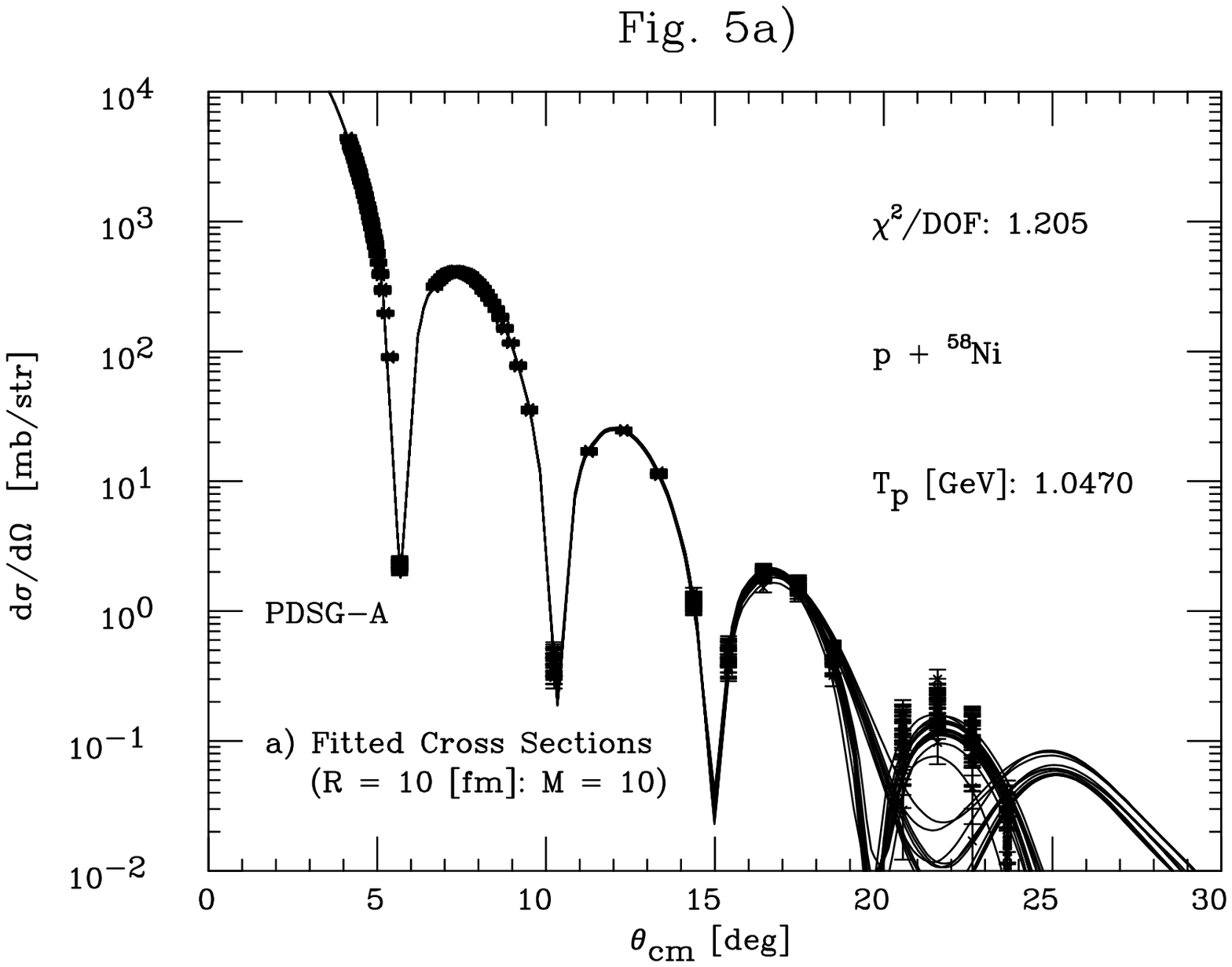}\\
    \includegraphics[width=30pc, keepaspectratio]{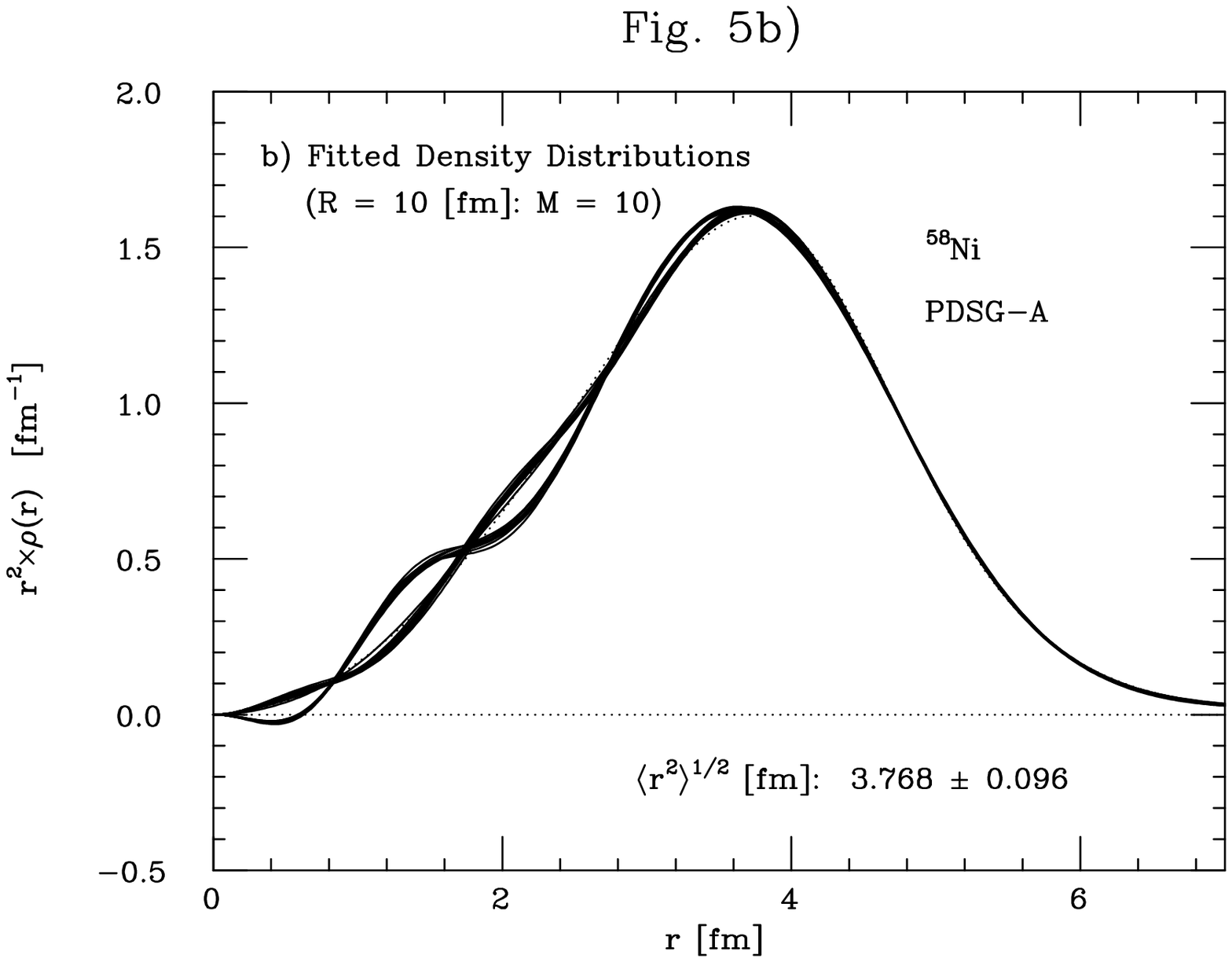}
  \end{center}
\caption{
Results of the least-square fitting for 
the pseudo-data-set group (PDSG) A of $^{58}$Ni.
$R$ $= 10$~[fm] and $M$ $= 10$. 
a) 25 fitted cross sections (solid curve) and all the pseudo-data (cross with bar). 
b) 25 fitted density distributions (solid curve), and the original distribution 
(dotted curve). The density distributions are drawn in the form of 
$r^2 \; \rho(r)$. 
} 
\end{figure}
\begin{figure}[h!]
  \begin{center}
    \includegraphics[width=30pc, keepaspectratio]{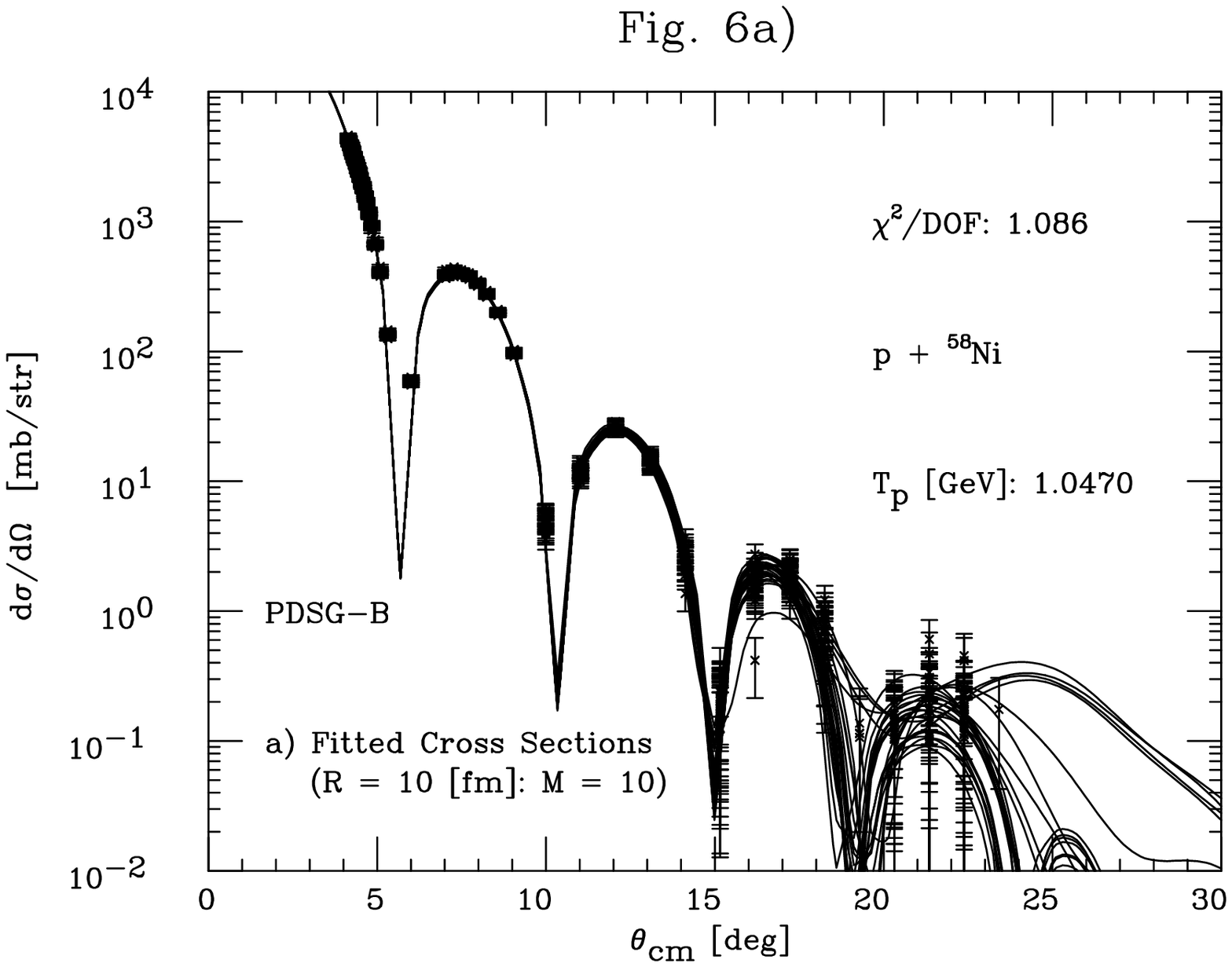}\\
    \includegraphics[width=30pc, keepaspectratio]{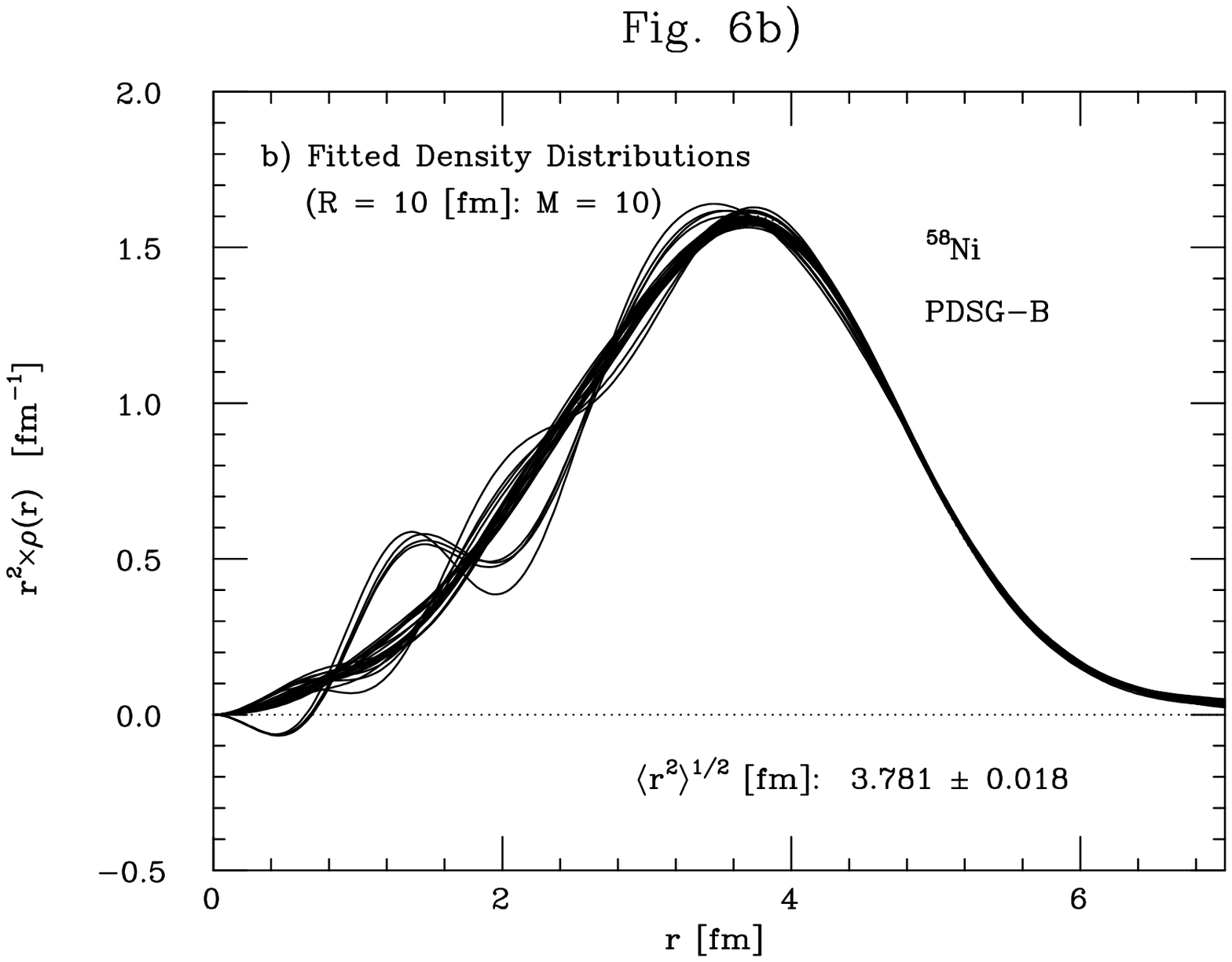}
  \end{center}
\caption{
Results of the least-square fitting for 
the pseudo-data-set group (PDSG) B of $^{58}$Ni.
$R$ $= 10$~[fm] and $M$ $= 10$. a) and b) are the same as Fig.~5. 
} 
\end{figure}
\begin{figure}[h!]
  \begin{center}
    \includegraphics[width=30pc, keepaspectratio]{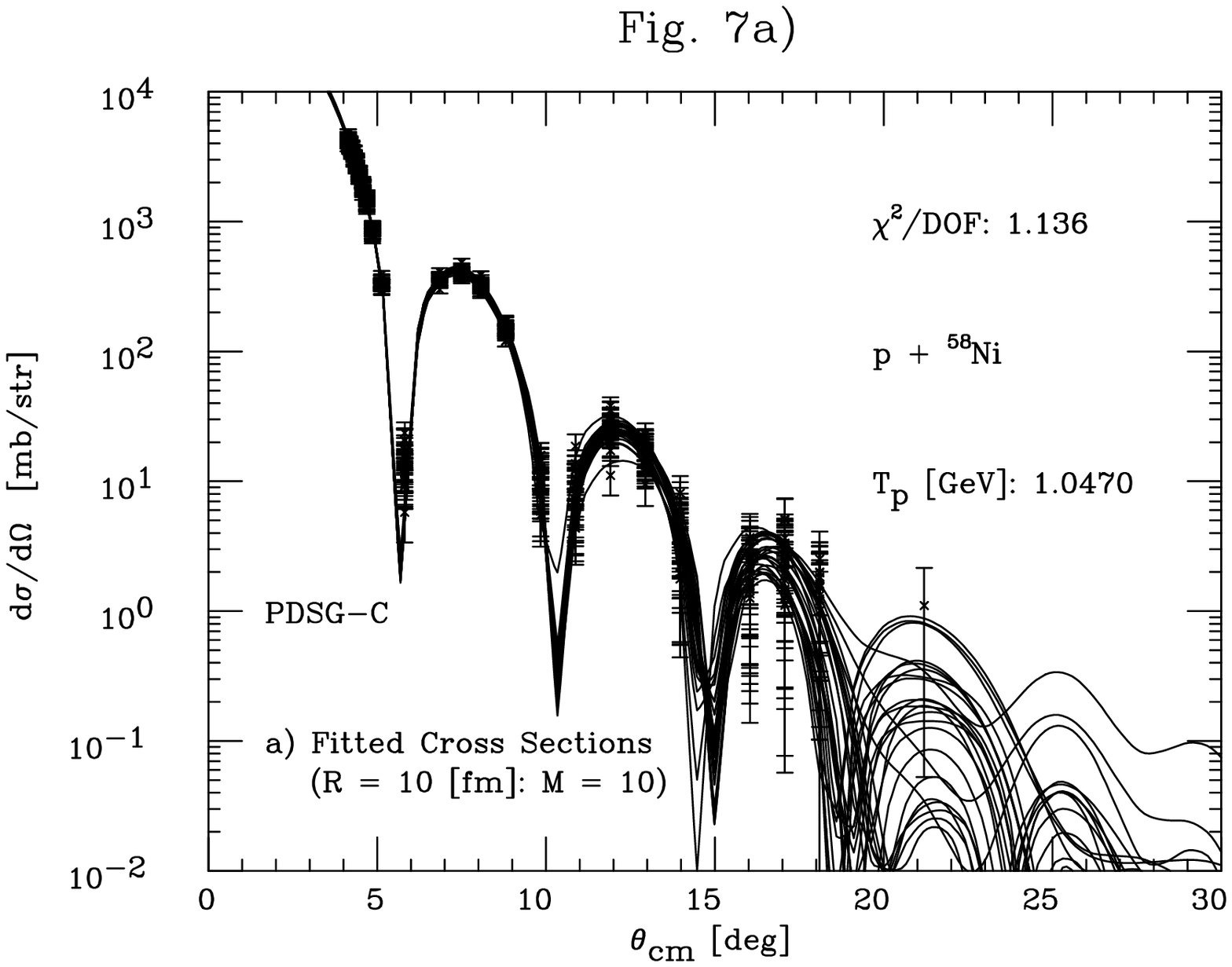}\\
    \includegraphics[width=30pc, keepaspectratio]{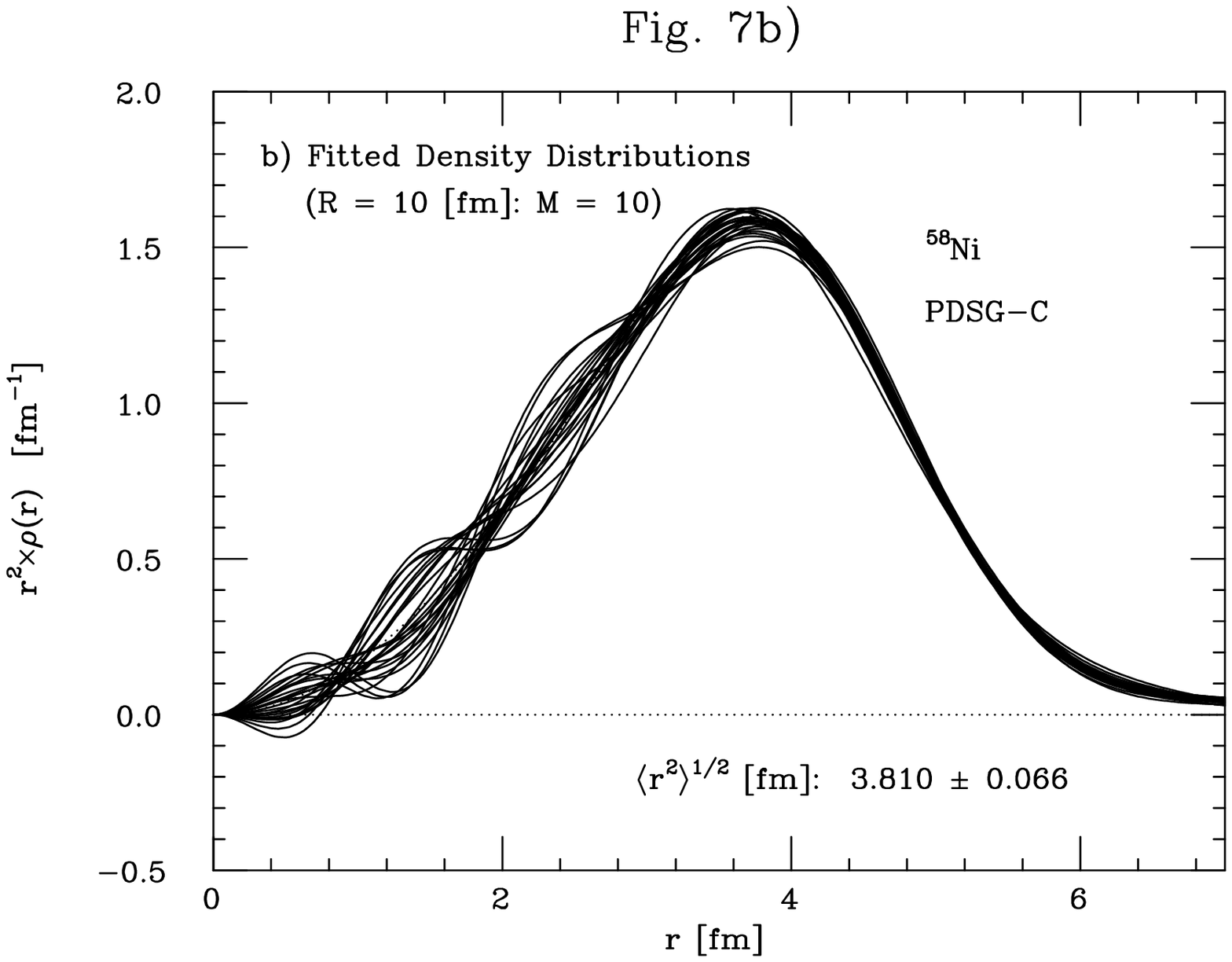}
  \end{center}
\caption{
Results of the least-square fitting for 
the pseudo-data-set group (PDSG) C of $^{58}$Ni.
$R$ $= 10$~[fm] and $M$ $= 10$. a) and b) are the same as Fig.~5. 
} 
\end{figure}
\begin{figure}[h!]
  \begin{center}
    \includegraphics[width=30pc, keepaspectratio]{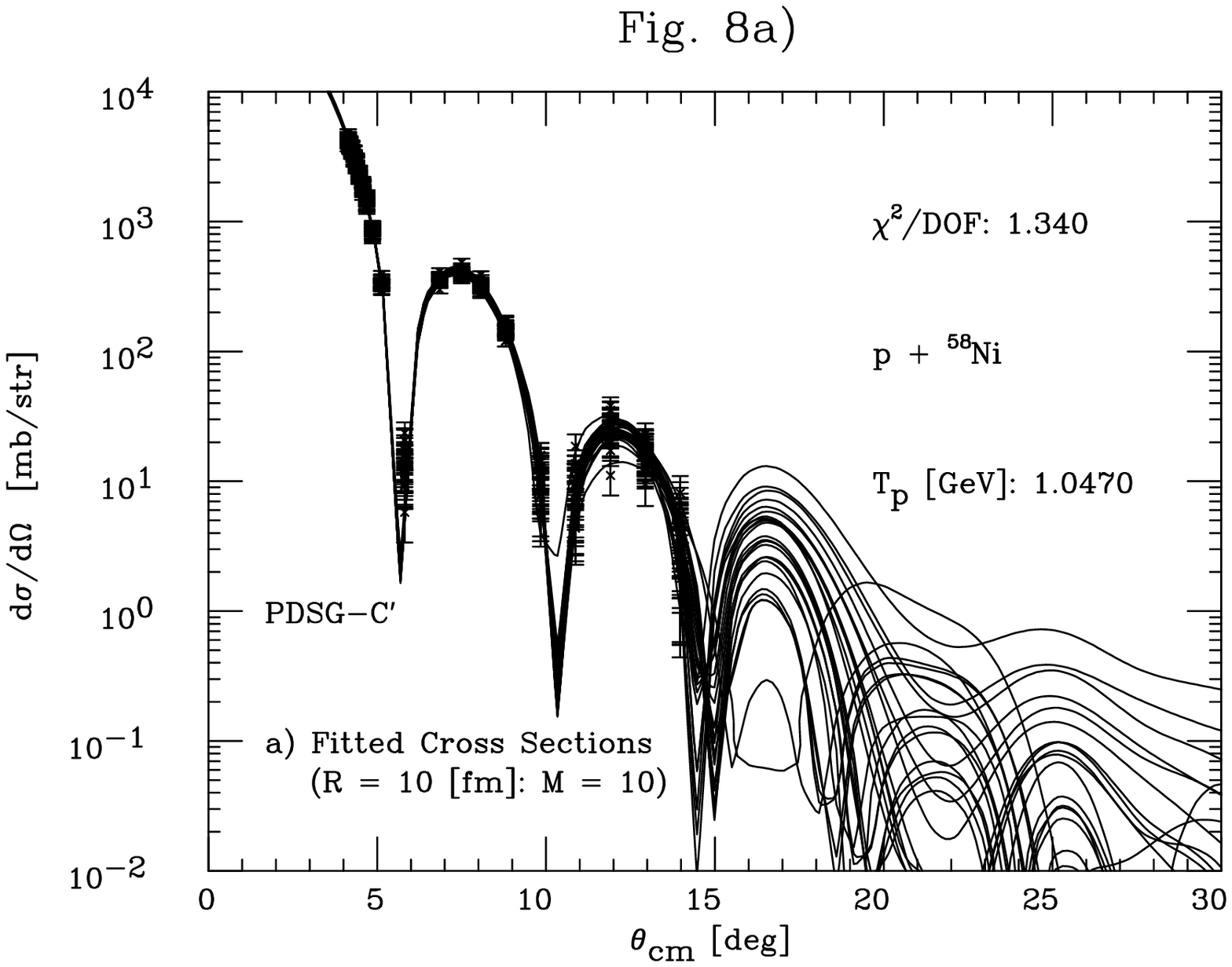}\\
    \includegraphics[width=30pc, keepaspectratio]{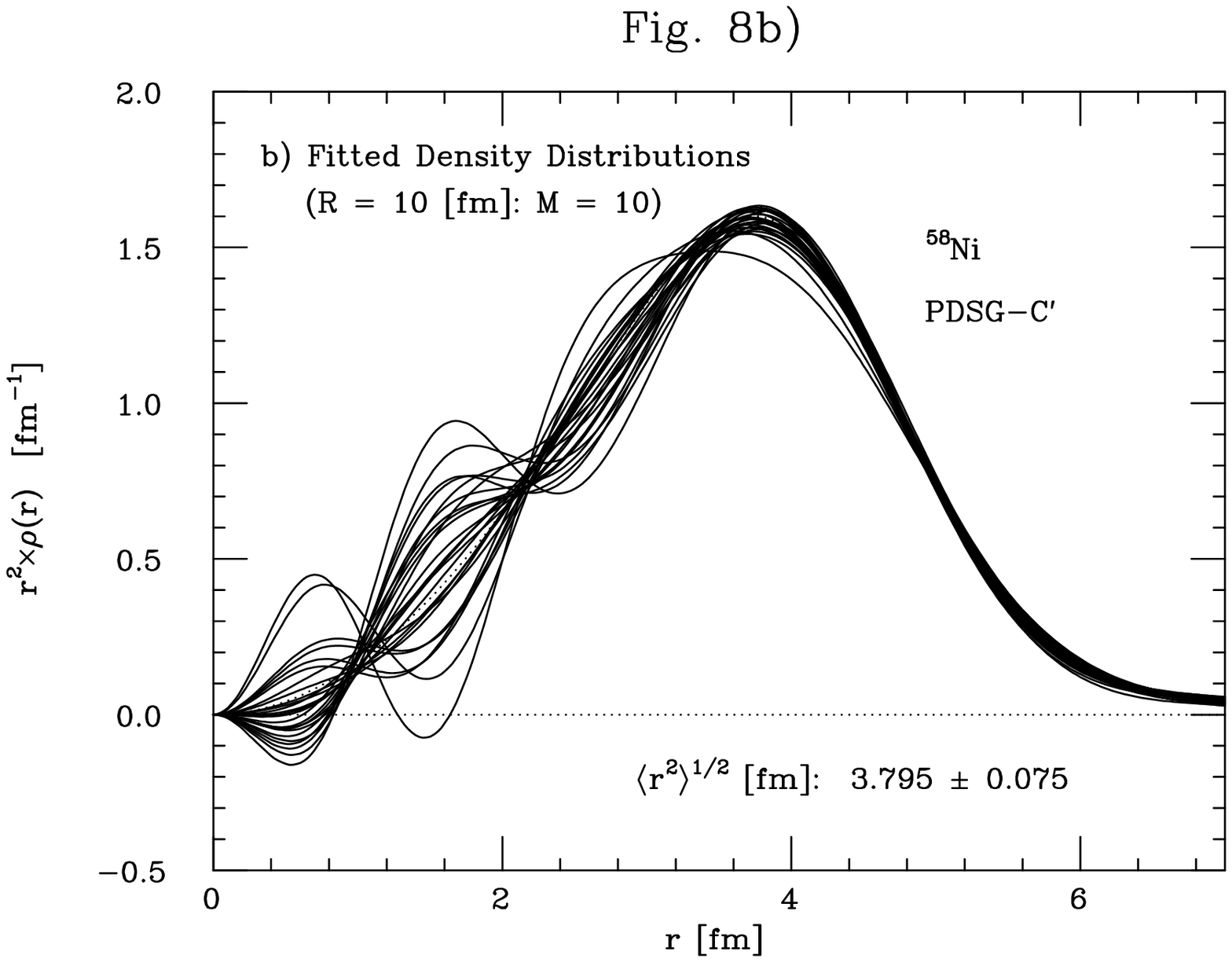}
  \end{center}
\caption{
Results of the least-square fitting for 
the pseudo-data-set group (PDSG) C' of $^{58}$Ni.
$R$ $= 10$~[fm] and $M$ $= 10$.. a) and b) are the same as Fig.~5. 
} 
\end{figure}
\begin{figure}[h!]
  \begin{center}
    \includegraphics[width=30pc, keepaspectratio]{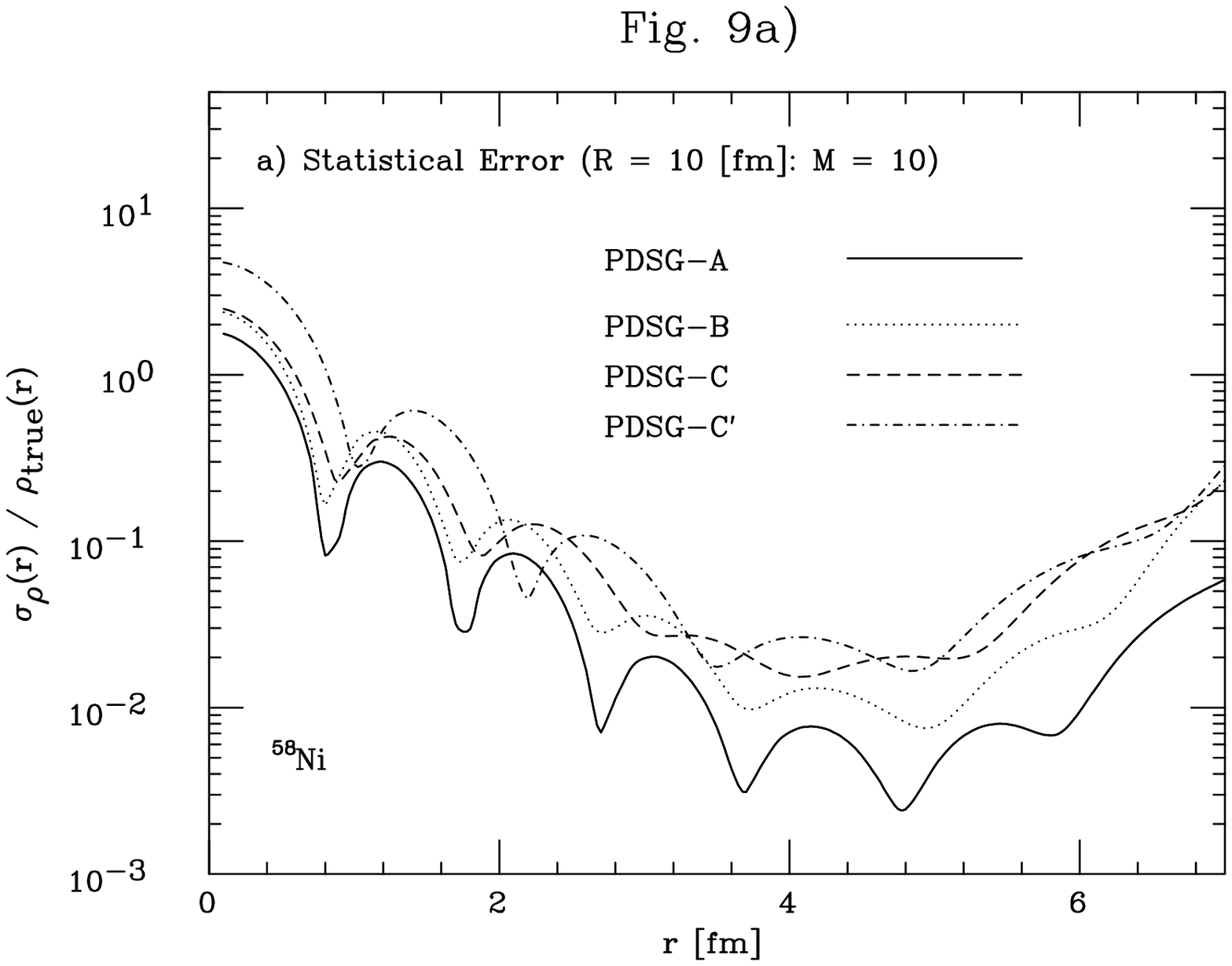}\\
    \includegraphics[width=30pc, keepaspectratio]{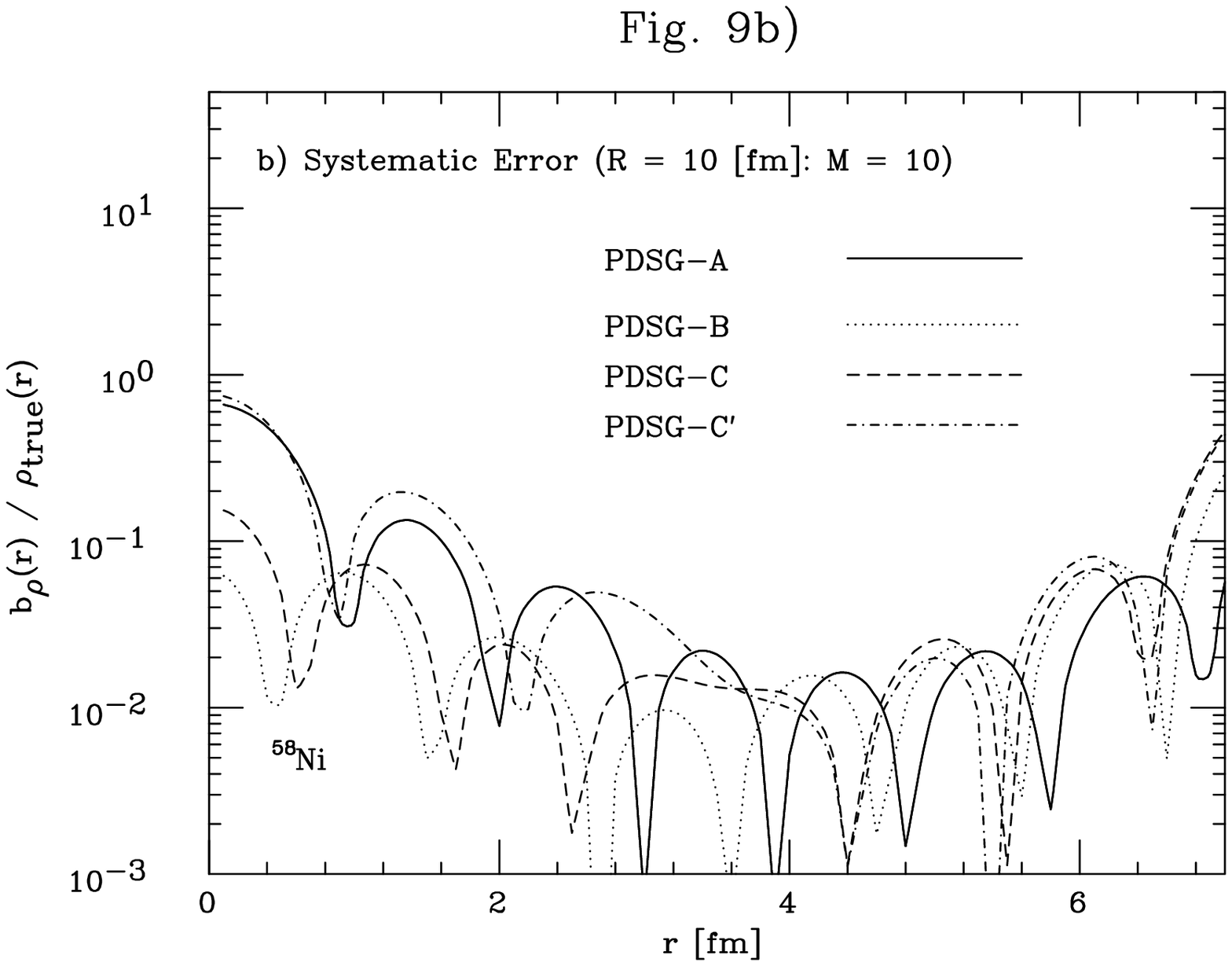}
  \end{center}
\caption{
a)~Statistical error, and b)~systematic error 
of the fitted density distributions for $^{58}$Ni. 
We divide them by $\rho_{\rm true}(r)$ at each $r$. 
The solid curve is for the pseudo-data-set group (PDSG) A, 
the dotted one is for the group B, 
the dashed one is for the group C,
the dash-dotted one is for the group C'. 
} 
\end{figure}
\begin{figure}[h!]
  \begin{center}
    \includegraphics[width=30pc, keepaspectratio]{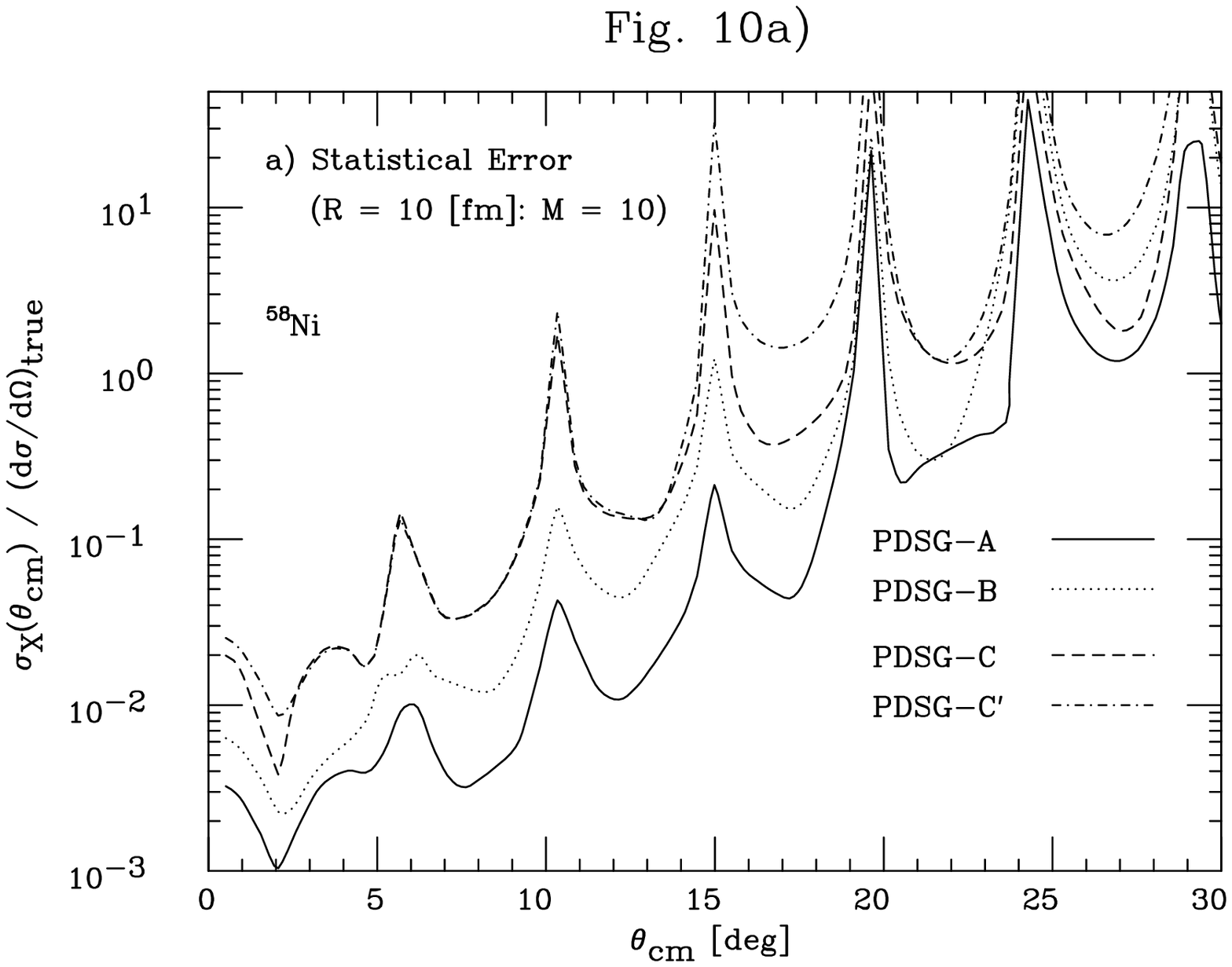}\\
    \includegraphics[width=30pc, keepaspectratio]{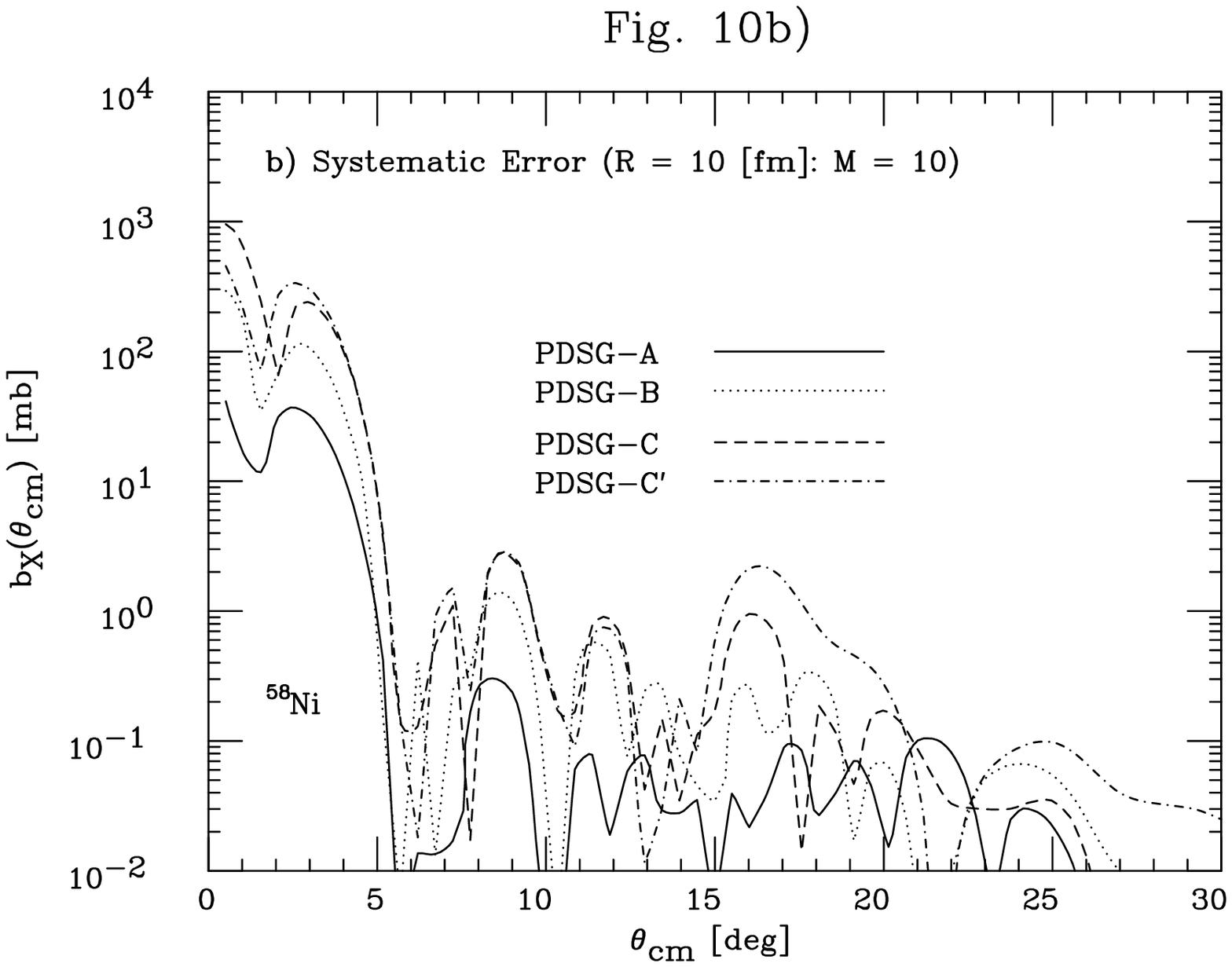}
  \end{center}
\caption{
a)~Statistical error, and b)~systematic error 
of the fitted cross sections for $^{58}$Ni. 
The curves are the same as in Fig.~9. 
} 
\end{figure}
\begin{figure}[h!]
  \begin{center}
    \includegraphics[width=30pc, keepaspectratio]{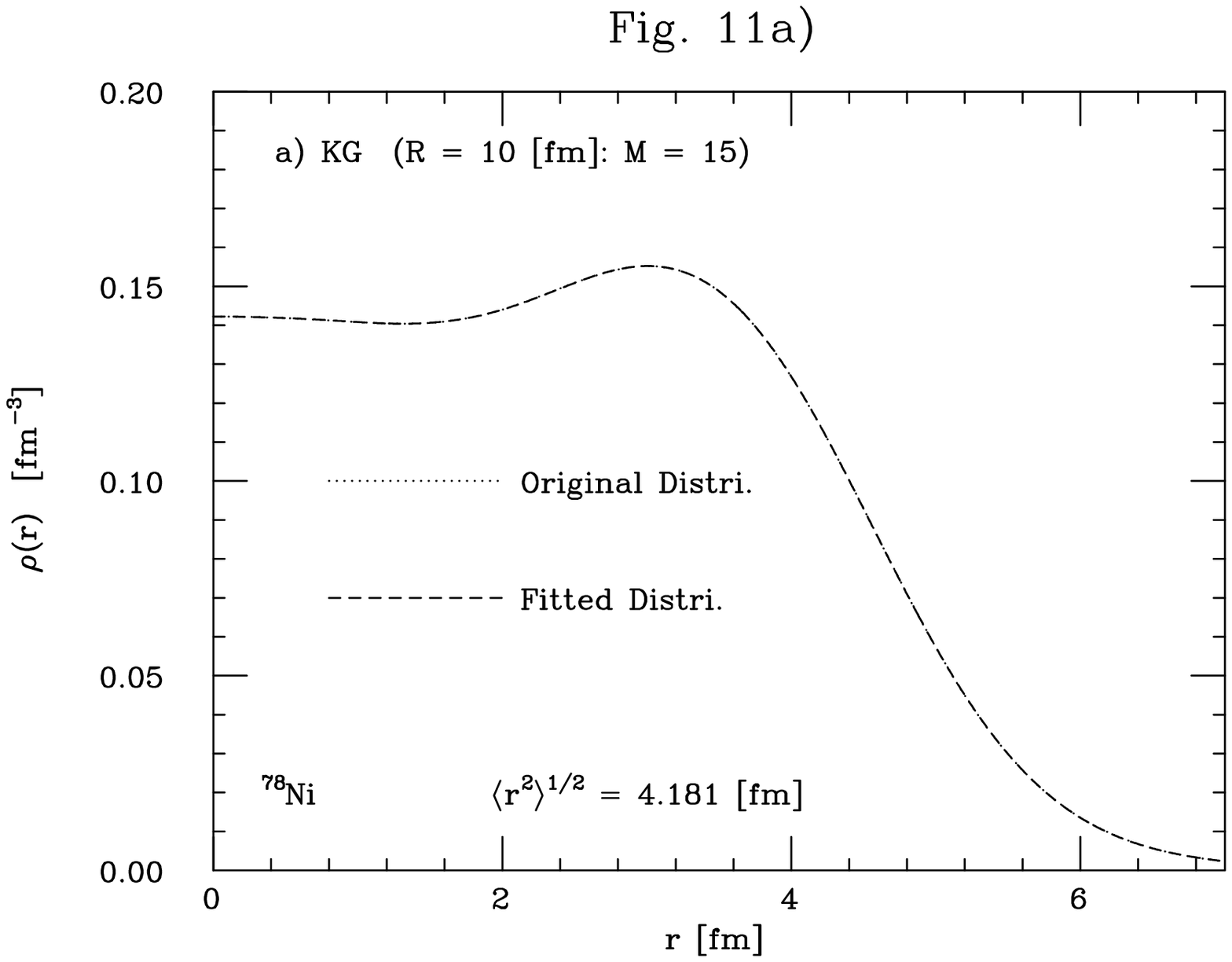}\\
    \includegraphics[width=30pc, keepaspectratio]{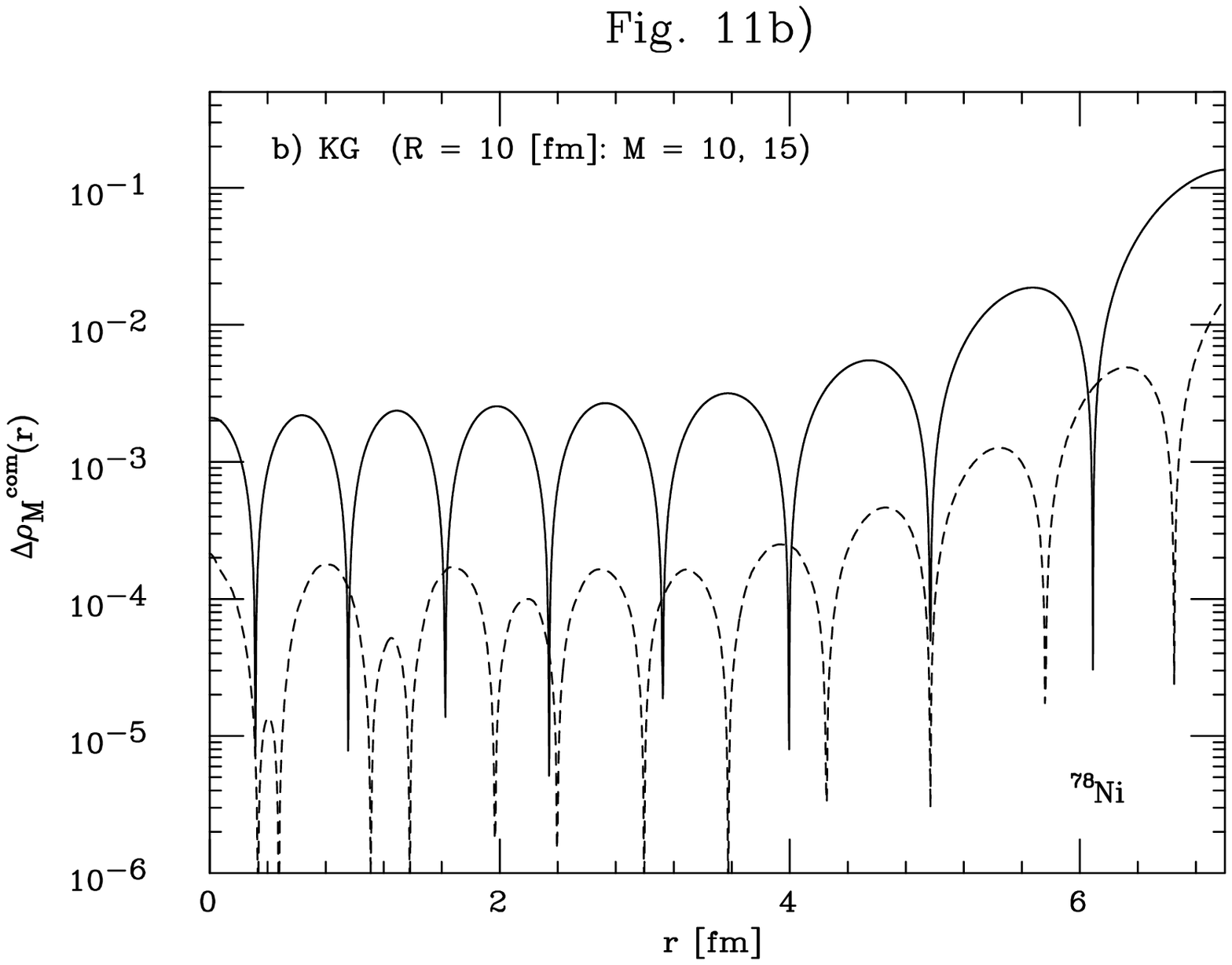}
  \end{center}
\caption{
a)~Matter density distribution of $^{78}$Ni. 
The dotted curve is the original density distribution 
obtained in the relativistic mean-field theory. $R$ $= 10$~[fm] and $M$ $= 15$.
The dashes curve is obtained by fitting 
of the Kamimura-Gauss (KG) basis functions. 
b)~Completeness error of $^{78}$Ni for $M$ $= 10$ and 15. 
The solid curve is for $M = 10$, 
$r_1$ $= 1.0$~[fm] and $r_M$ $= 5.5$~[fm] (Table~1).
The dashed curve is the case of $M = 15$,
$r_1$ $= 0.7$~[fm] and $r_M$ $= 5.6$~[fm] (Table~1).
} 
\end{figure}
\begin{figure}[h!]
  \begin{center}
    \includegraphics[width=30pc, keepaspectratio]{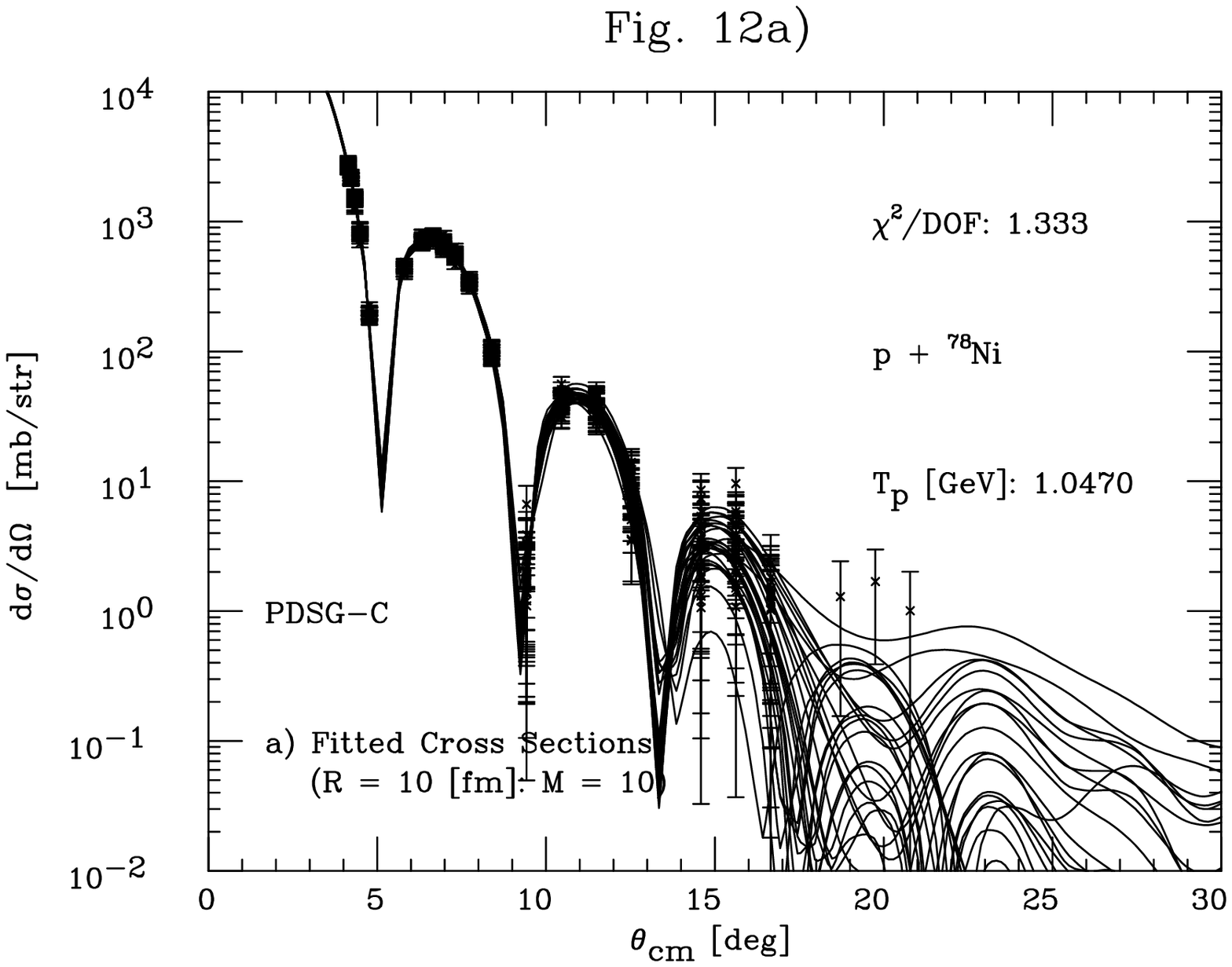}\\
    \includegraphics[width=30pc, keepaspectratio]{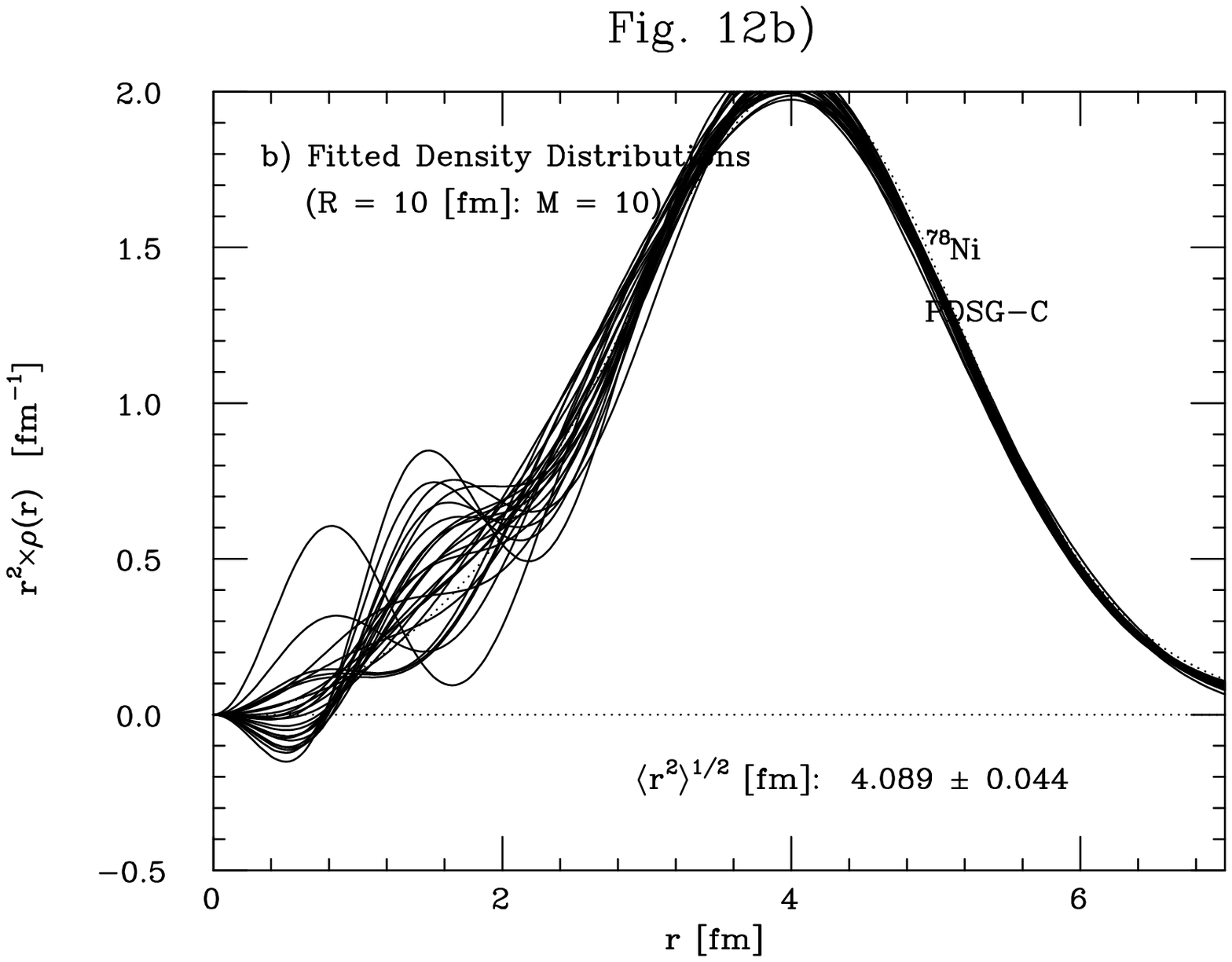}
  \end{center}
\caption{
Results of the least-square fitting 
for the pseudo-data-set group (PDSG) C of $^{78}$Ni.
$R$ $= 10$~[fm] and $M$ $= 10$. 
a) and b) are the same as Fig.~5. 
} 
\end{figure}
\begin{figure}[h!]
  \begin{center}
    \includegraphics[width=40pc, keepaspectratio]{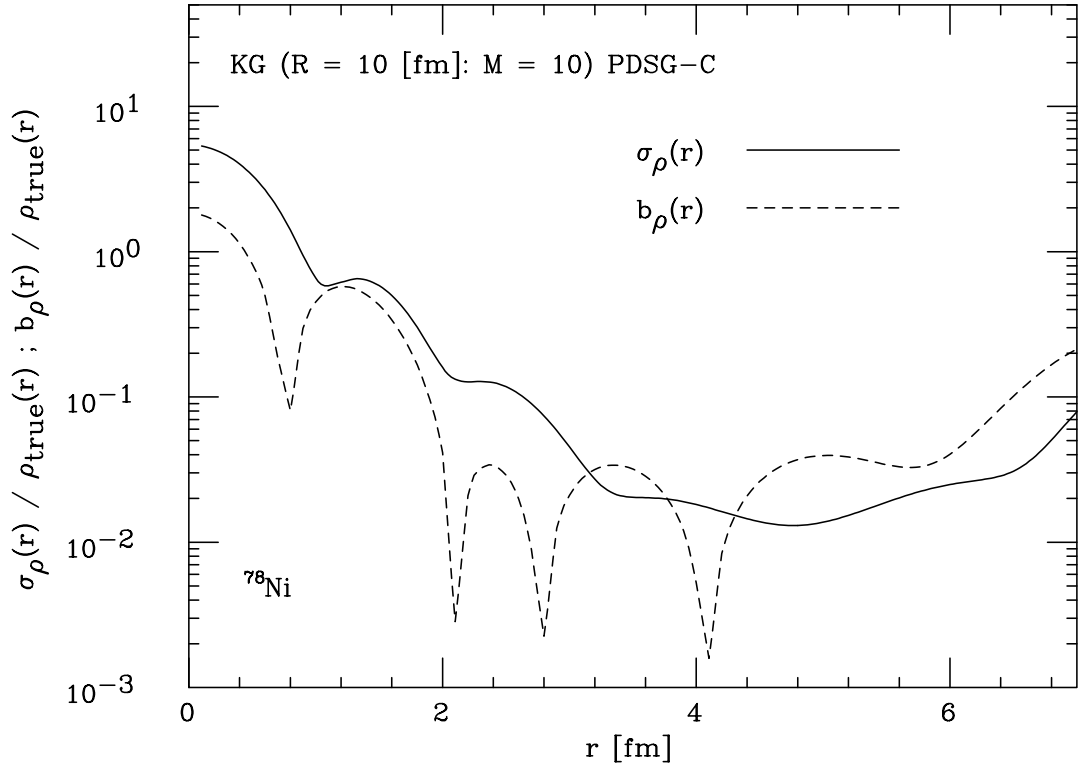}
  \end{center}
\caption{
Statistical error and systematic error 
of the fitted density distributions for $^{78}$Ni. 
The solid curve is the statistical error 
and the dashed one is the systematic error. 
They are divided by $\rho_{\rm true}(r)$. 
} 
\end{figure}
\begin{figure}[h!]
  \begin{center}
    \includegraphics[width=30pc, keepaspectratio]{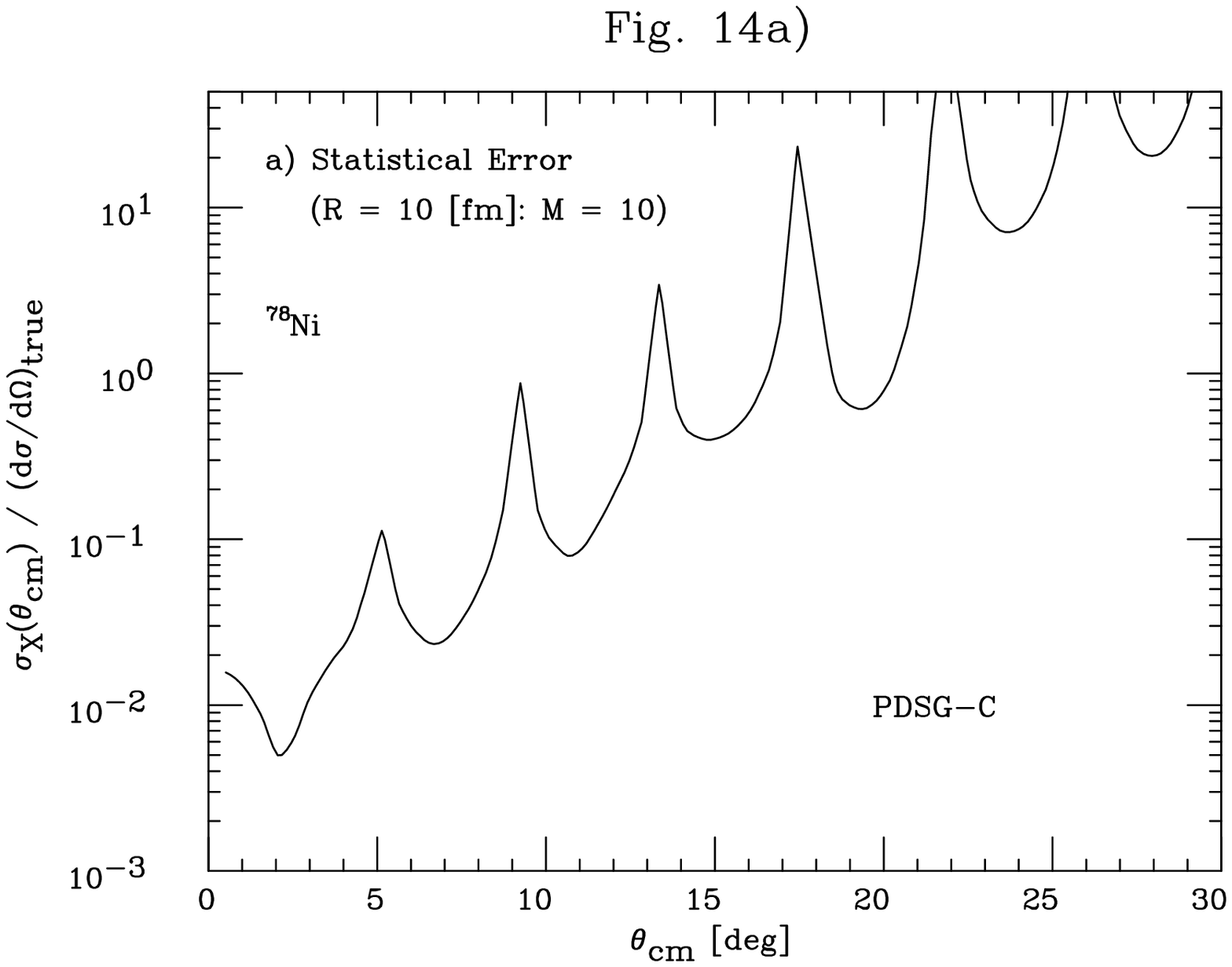}\\
    \includegraphics[width=30pc, keepaspectratio]{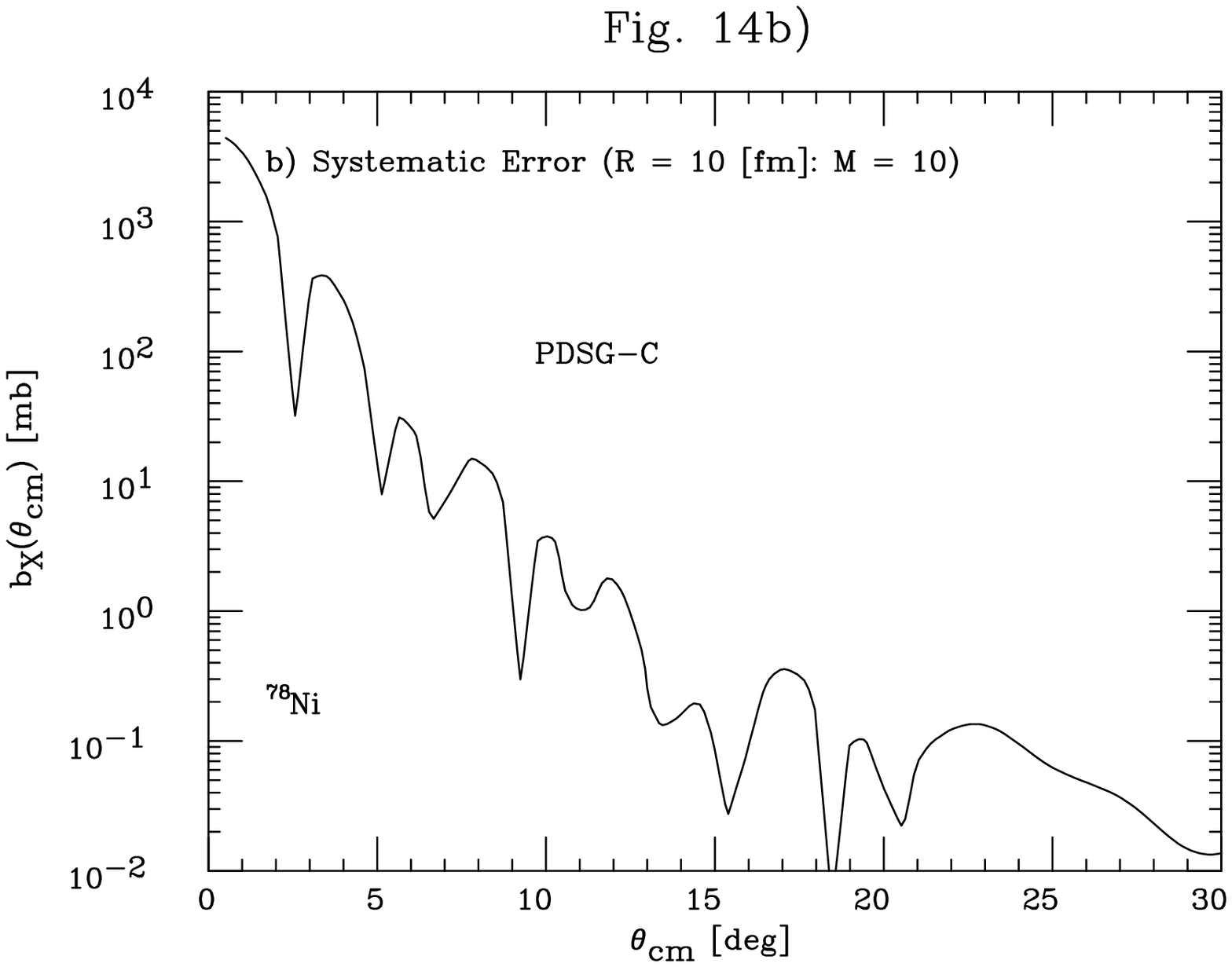}
  \end{center}
\caption{
a)~Statistical error, and b)~systematic error 
of the fitted cross sections for $^{78}$Ni. 
} 
\end{figure}
\begin{figure}[h!]
  \begin{center}
    \includegraphics[width=40pc, keepaspectratio]{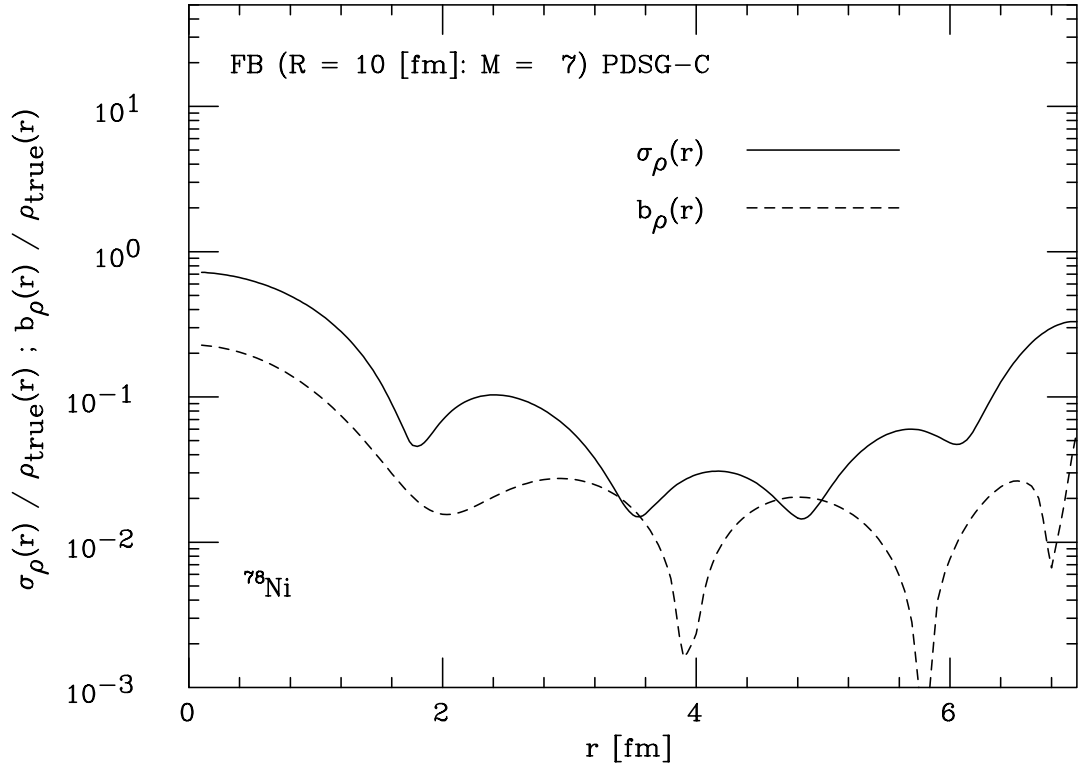}
  \end{center}
\caption{
Statistical error and systematic error 
of the fitted density distributions for $^{78}$Ni 
for the Fourier-Bessel (FB) basis functions. 
The curves are the same as in Fig.~13. 
} 
\end{figure}
\begin{figure}[h!]
  \begin{center}
    \includegraphics[width=30pc, keepaspectratio]{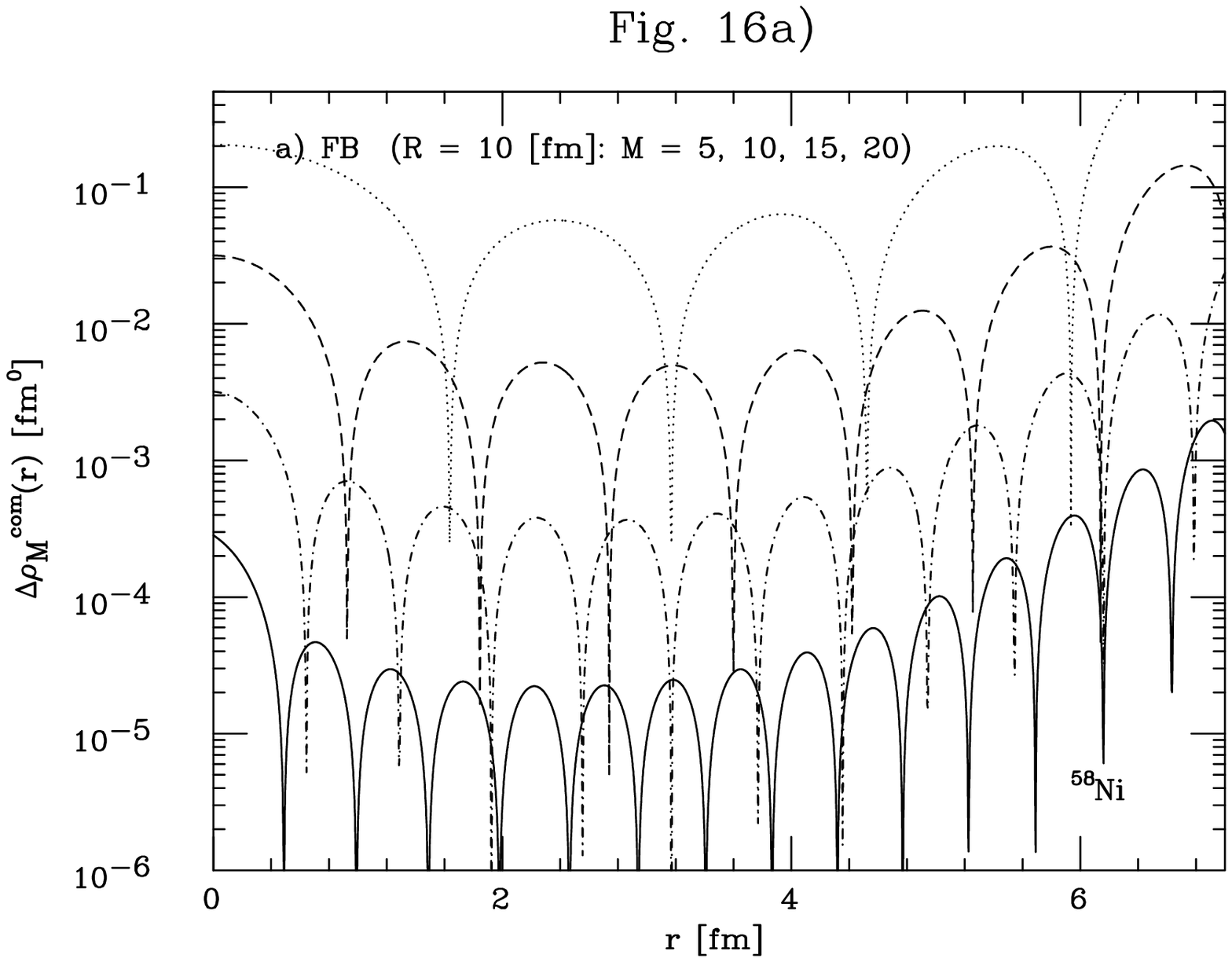}\\
    \includegraphics[width=30pc, keepaspectratio]{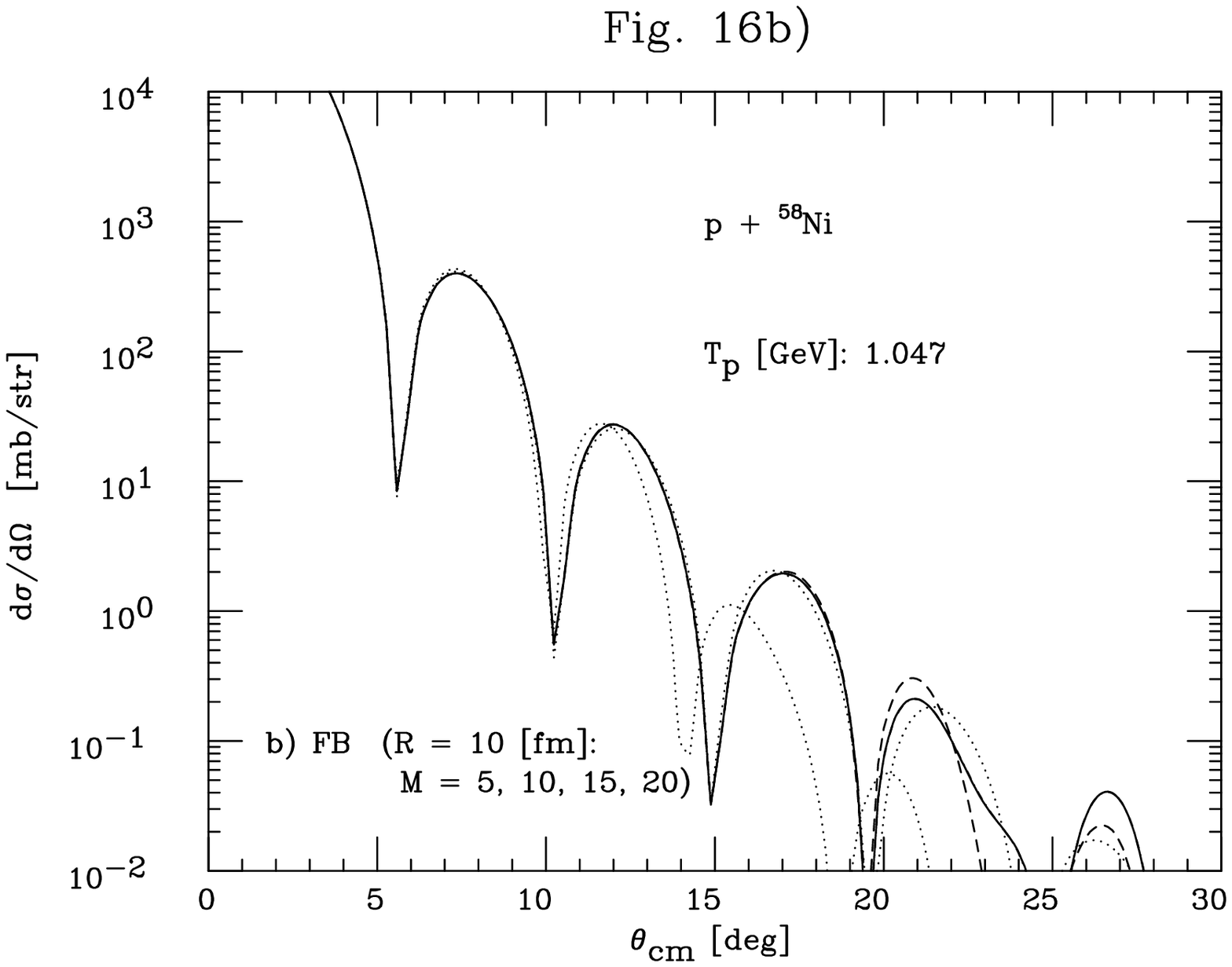}
  \end{center}
\caption{
$M$-dependence of a)~the completeness error 
of the Fourier-Bessel (FB) basis functions 
and of b)~the cross section with the same basis functions 
for $M =$ 5, 10, 15, and 20. $R$ $= 10$~[fm]. 
The dotted curve is for $M = 5$, 
the dashed one is for $M = 10$,
the dash-dotted one is for $M = 15$,
and the solid one is for $M = 20$.
Another dotted curve in b), as it is quite difficult to be seen, 
is obtained by using the original $\rho_{\rm true}(r)$. 
The curves are the same as in Figs.~2 and 3. 
} 
\end{figure}


\newpage
\begin{thebibliography}{99}
\bibitem{RIBF} Y.Yano, T.Katayama, A.Goto, and M.Kase, 
        {\em AIP Conf. Proc.}, {\bf 576}, (2001) 679. 
\bibitem{RIA} White Paper ``Scientific Opportunities with an Advanced 
         ISOL Facility", Nov., 1997. http://srfsrv.jlab.org/isol/
\bibitem{ANT:OX}A.N.Antonov, P.E.Hodgson, and I.Zh.Petkov, \\
        {\em Nucleon Momentum and Density Distributions in Nuclei}, \\
             Clarendon Press, 1988.
\bibitem{Tani} I.Tanihata {\it et al.}, 
        {\em Phys. Lett.}, {\bf B160}, (1985) 380; \\
        {\em Phys. Rev. Lett.}, {\bf 55}, (1985) 2676. 
\bibitem{Niigata} Proceedings of International Symposium on 
        {\em Structure and Reactions of Unstable Nuclei}, 
        ed.~K.Ikeda and Y.Suzuki, World Scientific, Singapore, (1991).
\bibitem{Ber:PR}C.A.Bertulani, L.F.Canto, and M.S.Hussein, 
        {\em Phys. Rep.}, {\bf 226}, (1993) 281.
\bibitem{Tani:JPG} I.Tanihata, {\em J. Phys.}, {\bf G22}, (1996) 157.
\bibitem{Elt} L.R.B.Elton, {\em Nuclear Sizes}, Oxford, (1961).
\bibitem{BM:NPA} D.Berdichevsky and U.Mosel, 
        {\em Nucl.Phys.}, {\bf A388}, (1982) 205.
\bibitem{Bat:ANP} C.J.Batty, E.Friedman, H.J.Gils, and H.Rebel,\\
        {\em Adv. Nucl. Phys.}, {\bf Vol.19}, (1989) 1.
\bibitem{Sic:LNP} I.Sick, {\em Lecture Notes in Physics}, {\bf 236}, (1985) 137.
\bibitem{Fro:MT} B.Frois, C.N.Papanicolas, and S.E.Williamson, 
        {\em Modern Topics in Electron Scattering}, 
        ed.~B.Frois and I.Sick, World Scientific, (1991) 352.  
\bibitem{FN:NP} J.L.Friar and J.W.Negele,
        {\em Nucl. Phys.}, {\bf A212}, (1973) 93; \\
        {\em Adv. Nucl. Phys.}, {\bf Vol.8}, (1975) 219.
\bibitem{Cowan} G.Cowan, {\em Statistical Data Analysis}, 
        Clarendon Press, Oxford, (1998).
\bibitem{MEM1} {\em Maximum-Entropy and Bayesian Methods 
        in Science and Engineering}, Vol.~1:~Foundations. 
        ed.~G.J.Erickson and C.Ray Smith, Kluwer Academic Publishers, 
        (1988).
\bibitem{drip} A.Kohama, R.Seki, A.Arima, and S.Yamaji, 
        {\em RIKEN Review}, {\bf 39}, (2001) 155.
\bibitem{Glau:Lec}R.J.Glauber,
        {\em Lectures in Theoretical Physics}, \\
        ed.~W.E.Brittin and D.G.Dunham, 
        Interscience, New York, {\bf Vol.1}, (1959) 315;
        {\em High-Energy Physics and Nuclear Structure}, 
        ed.~ S.Devons, Plenum Press (1970) 207.
\bibitem{Dre:NPA} B.Dreher, J.Friedrich, K.Merle, H.Rothhaus, and G.Luehrs,\\
        {\em Nucl. Phys.}, {\bf A235}, (1974) 219.
\bibitem{FV:NPA} J.Friedrich and N.Voegler,
        {\em Nucl. Phys.}, {\bf A373}, (1982) 192.
\bibitem{FVR:NPA} J.Friedrich, N.Voegler, and P.-G.Reinhard, 
        {\em Nucl. Phys.}, {\bf A459}, (1986) 10.
\bibitem{Kam:PRA} M.Kamimura, {\em Phys. Rev.}, {\bf A38}, (1988) 621..
\bibitem{Lenz:ZP} F.Lenz,  {\em Z.Phys.}, {\bf 222}, (1969) 491.
\bibitem{FL:NP} J.Friedrich and F.Lenz, 
        {\em Nucl. Phys.}, {\bf A183}, (1972) 523.
\bibitem{Sic:NP} I.Sick, {\em Nucl. Phys.}, {\bf A218}, (1974) 509.
\bibitem{Bre:PRL} V.Breton {\it et al.}, 
        {\em Phys. Rev. Lett.}, {\bf 66}, (1991) 572. 
\bibitem{Cave:PRL} J.M.Cavedon {\it et al.},
        {\em Phys. Rev. Lett.}, {\bf 58}, (1987) 195.
\bibitem{Eld:SJNP} Yu.N.Eldyshev, V.N.Lukyanov, Yu.S.Pol', 
        {\em Sov. Jour. Nucl. Phys.}, {\bf 16}, (1973) 282.
\bibitem{Bur:PAN} V.V.Burov, D.N.Kadrev, V.K.Lukyanov, and Yu.S.Pol',\\
        {\em Phys. Atom. Nucl.}, {\bf 61}, (1998) 525.
\bibitem{Fic:PLB} J.R.Ficenec, W.P.Trower, J.Heisenberg, and I.Sick,
        {\em Phys. Lett.}, {\bf 32B}, (1970) 460.
\bibitem{Sic:PRL35} I.Sick {\it et al.}, 
        {\em Phys. Rev. Lett.}, {\bf 35}, (1975) 910. 
\bibitem{Alk:PLB} G.D.Alkhazov {\it et al.}, 
        {\em Phys. Lett.}, {\bf 67B}, (1977) 402-404.
\bibitem{Alk:NPA} G.D.Alkhazov {\it et al.}, 
        {\em Nucl. Phys.}, {\bf A381}, (1982) 430. 
\bibitem{Lom:NPA} R.M.Lombard, G.D.Alkhazov, and O.A.Domchenkov, \\
        {\em Nucl. Phys.}, {\bf A360}, (1981) 233. 
\bibitem{ADL80} R.D.Amado, J.P.Dedonder, F.Lenz, 
        {\em Phys. Rev.}, {\bf C21}, (1980) 647. 
\bibitem{AWT:NP81} A.W.Thomas, {\em Nucl. Phys.}, {\bf A354}, (1981) 51c.
\bibitem{Alk:PR} G.D.Alkhazov, S.L.Belostotsky, and A.A.Vorobyov,
        {\em Phys. Rep.}, {\bf 42C}, (1978) 89.
\bibitem{Cha:AP} A.Chaumeaux, V.Layly, and R.Schaeffer,
        {\em Ann. Phys.}, {\bf 116}, (1978) 247.
\bibitem{PRC:Bri} I.Brissaud and M.K.Brussel,
        {\em Phys. Rev.}, {\bf C15}, (1977) 452.
\bibitem{Ray:PRC} L.Ray, W.Rory Coker, and G.W.Hoffmann,
        {\em Phys. Rev.}, {\bf C18}, (1978) 2641. 
\bibitem{Ray:PRC79} L.Ray, {\em Phys. Rev.}, {\bf C19}, (1979) 1855.
\bibitem{Hof:PRC} G.W.Hoffmann {\it et al.},
        {\em Phys. Rev.}, {\bf C21}, (1980) 1488.
\bibitem{Str:PRC} V.E.Starodubsky and N.M.Hintz,
        {\em Phys. Rev.}, {\bf C49}, (1994) 2118.
\bibitem{Gil:PRC} H.J.Gils, H.Rebel, and E.Friedman, 
        {\em Phys. Rev.}, {\bf C29}, (1984) 1295.
\bibitem{Alk:PRL} G.D.Alkhazov {\it et al.}, 
        {\em Phys. Rev. Lett.}, {\bf 78}, (1997) 2313.
\bibitem{seki} R.Seki and I.Tanihata, unpublished. 
\bibitem{Mar:PRC} F.Mar\'{e}chal {\it et al.}, 
        {\em Phys. Rev.}, {\bf C60}, (1999) 034615.
\bibitem{Sch:PRC} H.Scheit {\it et al.}, 
        {\em Phys. Rev.}, {\bf C63}, (2000) 014604.
\bibitem{Tani:PL} I.Tanihata {\it et al.}, 
        {\em Phys. Lett.}, {\bf B287}, (1992) 307; 
        {\em Phys. Lett.}, {\bf B289}, (1992) 261. 
\bibitem{Ozawa1} A.Ozawa {\it et al.}, 
        {\em Nucl.Phys.}, {\bf A691}, (2001) 599.
\bibitem{Ozawa} A.Ozawa, T.Suzuki, and I.Tanihata, 
        {\em Nucl. Phys.}, {\bf A693}, (2001) 32.
\bibitem{minuit} F.James, MINUIT, Reference Manual, Ver.~94.1, 
         CERN Program Library Long Writeup {\bf D506}, (1994).
\bibitem{GM:NPB} R.Glauber and G.Matthiae, 
        {\em Nucl. Phys.}, {\bf B21}, (1970) 135.
\bibitem{czy:NPB} W.Czy\.{z}, L.Le\'{s}niak, and W.Wo\mbox{\l}ek,
        {\em Nucl. Phys.}, {\bf B19}, (1970) 125.
\bibitem{DeV:ATO} H.DeVries, W.DeIager, and C.DeVries, \\
        {\em Atomic Data and Nuclear Data Tables}, {\bf 36}, (1987) 495. 
\bibitem{BH:PRC} J.Borysowicz and J.H.Hetherington, 
        {\em Phys.Rev.}, {\bf C7}, (1973) 2293.
\bibitem{HB:NP} J.H.Hetherington and J.Borysowicz, 
        {\em Nucl. Phys.}, {\bf A219}, (1974) 221.
\bibitem{Lit:PRC} J.W.Lightbody, Jr. {\it et al.}, 
        {\em Phys.Rev.}, {\bf C27}, (1983) 113.
\bibitem{Men:PLB} J.Meng, I.Tanihata, and S.Yamaji, 
        {\em Phys.Lett.}, {\bf B419}, (1998) 1.
\bibitem{Scl:PLB} S.Shlomo and R.Schaeffer, 
        {\em Phys. Lett.}, {\bf 83B}, (1979) 5.
\bibitem{Var:pre} K.Varga, S.C.Pieper, Y.Suzuki, and R.B.Wiringa, 
        {\em nucl-th/0205027}.
\bibitem{Hi:NP}E.Hiyama and M.Kamimura,
        {\em Nucl. Phys.}, {\bf A588}, (1995) 35c.
\bibitem{LB} J.Bystricky {\it et al.}, 
        {\em LANDOLT-B\"{O}RNSTEIN. New Series.}, I/9a (1980).
\end{thebibliography}
\end{document}